\documentclass{aa}  

\usepackage{graphicx}
\usepackage{amsmath}

\usepackage{graphicx}
\usepackage{txfonts}
\usepackage[hidelinks]{hyperref}
\usepackage[usenames,dvipsnames]{color}
\usepackage{tablefootnote}

\makeatletter
\renewcommand*\aa@pageof{,page \thepage{} of \pageref*{LastPage}}

\begin{document}

   \title{PEWDD: A database of white dwarfs enriched by \\exo-planetary material}

   \author{J. T. Williams
          \inst{1},
          B. T. G{\"a}nsicke\inst{1,2},
          A. Swan\inst{1},
          M. W. O'Brien\inst{1},
          P. Izquierdo\inst{1},
          A-M. Cutolo\inst{1}\\
          \and
          T. Cunningham\inst{3}\thanks{NASA Hubble Fellow}
          }

   \institute{Department of Physics, University of Warwick, Coventry, CV4 7AL, UK\\
              \email{jamietwilliams.astro@gmail.com}
         \and
             Centre for exo-planets and Habitability, University of Warwick, Coventry CV4 7AL, UK
         \and
             Center for Astrophysics | Harvard \& Smithsonian, 60 Garden St., Cambridge, MA 02138, USA
             }

   \date{Accepted September 23, 2024}

 
  \abstract
   {We present the Planetary Enriched White Dwarf Database (PEWDD), a collection of published photospheric abundances of white dwarfs accreting planetary debris alongside additional information relevant to metal-enrichment and the presence of infrared excesses, emission lines, and binary companions. PEWDD contains at the time of publishing information on 1739 white dwarfs and will be kept up-to-date with information from new publications. A total of 24 photospheric metals are recorded and are linked to accretion of exo-planetary material. The overall properties of metal-enriched white dwarfs are severely affected by observational selection effects; in particular we find that what metals are detectable strongly correlates with the effective temperature. By considering metal-enriched white dwarfs which have abundances measured by different methods, we find a spread that is comparable with the often quoted ad-hoc estimated abundance uncertainties, i.e. $\simeq0.1-0.2$\,dex. We draw attention to a dichotomy in the median accretion rates for metal-enriched H- and He-dominated white dwarfs, with $\dot M_{\text{acc, H}} = 7.6 \times 10^7\,\mathrm{g\,s^{-1}}$ and $\dot M_{\text{acc, He}} = 8.7 \times 10^8\,\mathrm{g\,s^{-1}}$ when extrapolating bulk compositions from bulk Earth Ca abundance. We identify 40 metal-enriched white dwarfs in binary systems and find evidence that enrichment is suppressed by binary companions within 200\,au.}

   \keywords{planetary systems -- white dwarfs -- stars: abundances}
    \titlerunning{Planetary enriched white dwarf database}
    \authorrunning{J. Williams et al.}
   \maketitle
%

\section{Introduction}

The final evolutionary stage of 97~per cent of main sequence stars is a white dwarf: an electron degenerate star with such a high surface gravity that all elements heavier than He, from now on metals, rapidly diffuse downwards, leaving a purely H- or He-dominated atmosphere \citep{FontaineMichaud1979}. However, many white dwarfs are enriched with traces of photospheric metals in their atmosphere \citep{vanMaanen1917,Zuckerman2003,Zuckerman2010,Koester2014_DAZ,Hollands2017}. This metal enrichment posed a serious puzzle, since the abundances of metals were too high to be explained by accretion from the interstellar medium  \citep{Koester1976,Wesemael1979,Dupuis1993} or radiative levitation alone \citep{Chayer1995}. An infrared excess discovered at the white dwarf G\,29-38 \citep{Zuckerman1987} led to the realisation that the metal-enrichment arises from the disruption, and subsequent accretion of asteroids \citep{Jura2003}. The identification of many additional circumstellar debris discs around other metal-enriched white dwarfs corroborated that scenario \citep{Becklin2005, Kilic2007, Jura2007, FarihiDiscs2009}.

Exo-planets or exo-asteroids can survive the red giant branch phase of stellar evolution if located far enough away from the star and there are dynamical pathways through which they can be scattered towards the white dwarf \citep{AntoniadouVeras2016,Mustill2018,Veras2022}. Once within the Roche radius, the asteroids are disrupted by tidal forces, forming a circumstellar disc and material is then accreted by the white dwarf \citep{BochkarevRafikov2011}. Material impacting the white dwarf surface has been directly observed using X-rays \citep{Cunningham2022_xray}. Metal-enrichment indicating planetary accretion implies that the discovery of photospheric Ca in the spectrum of the white dwarf van Maanen~2 in 1917 was the first trace of an exo-planetary system \citep{Zuckerman2015}.

Analysis of metal-enriched white dwarfs can provide detailed information on the bulk composition of exo-planets that is unobtainable by any other method. Measuring parent body compositions was first pioneered in \citet{Zuckerman2007}, using state-of-the-art atmospheric models \citep[e.g.][]{Koester2009_WDs,Dufour2010,Blouin2018a} to measure the photospheric metal abundances from the analysis of optical or ultraviolet (UV) spectroscopy. By considering the equilibrium between the downward diffusion flux and accretion rate into the white dwarf atmosphere, the chemical composition of accreted material can be determined \citep{Koester2009_WDs}. A large diversity of exo-planetary material has been identified, ranging from material resembling bulk Earth or chondrites \citep[e.g.][]{Izquierdo2021,Doyle2023,Trierweiler2023}, material being rich in core-like \citep[e.g.][]{Melis2011,Wilson2015}, mantle-like \citep[e.g.][]{Zuckerman2011,MelisDufour2017}, or crust-like material \citep[e.g.][]{Hollands2021}, hydrated asteroids \citep[e.g.][]{Farihi2013,GentileFusillo2017,Hoskin2020}, icy objects \citep[e.g.][]{Xu2017,Klein2021,Doyle2021} and even the atmosphere from an evaporating giant planet \citep[e.g.][]{Gaensicke2019}.

In this paper, we introduce the Planetary Enriched White Dwarf Database (hereafter PEWDD) which gathers information about white dwarfs that were identified to be metal-enriched by planetary material from the literature. Unlike previous large-scale studies of metal-enriched white dwarfs that were focused either on a single element, (e.g. Ca in \citealt{Coutu2019}) or very specific samples of white dwarfs \citep[e.g.][]{Koester2014_DAZ, Hollands2017,Hollands2018}, PEWDD contains the published information on all detected metals across the entire range where the accretion of planetary debris is the currently preferred interpretation for a given system. This is a heterogeneous sample, biased towards brighter white dwarfs, where observations with higher spectral resolution and greater signal-to-noise ratio (S/N) facilitate the detection of multiple elements. Nevertheless, this collection of the abundances of exo-planetary material already enables powerful statistical analyses, and its scientific potential will grow as additional systems are discovered and analysed.

The information available in PEWDD and a discussion on how to make use of it are given in Section \ref{sec:structure}. Properties of metal-enriched white dwarfs and potential observational biases are explored in Sect. \ref{sec:generic}. We provide code in order to create number abundance and mass fraction diagrams using PEWDD and examples are given in Sect. \ref{sec:plots}. The spread in number abundances stemming from the use of different methods and observational data is investigated in Sect. \ref{sec:errors}. In Sect. \ref{sec:accretion} we discuss the accretion rate dichotomy, mass of accreted material and discs. In Sect. \ref{sec:magnetic} we discuss magnetic metal-enriched white dwarfs and in Sect. \ref{sec:binaries} we examine binaries. Conclusions are provided in Sect. \ref{sec:conclusion}.

\section{Database structure}\label{sec:structure}

To be included in PEWDD, a white dwarf needs to have published photospheric abundances\footnote{Unless explicitly mentioned, abundances are expressed by number throughout the paper.} which is most likely the result of ongoing or recent accretion of planetary material\footnote{We hence exclude white dwarfs with photospheric metals caused purely by radiative levitation with no evidence of recent accretion \citep{Chayer1995}, dredge-up \citep{Koester2020,Bedard2022}; extremely hot white dwarfs which have not yet gravitationally settled and hence exhibit metal-rich atmospheres \citep{Werner2017, Werner2018}, accretion from winds of close stellar or sub-stellar companions \citep[e.g.][]{ODonoghue2003, Debes2006, Tappert2011, Pyrzas2012,Walters2023} or the very metal-rich partially burned remnants of thermonuclear supernovae \citep{Vennes2017, Raddi2019, Gaensicke2020}.}. At the time of publication, PEWDD contains 3501 unique abundance measurements of 1739 individual white dwarfs. There are more entries than white dwarfs due to cases where multiple papers study the same white dwarf, or papers that contain multiple abundance measurements using different parameters for a single white dwarf. Where available, we include additional physical parameters, such as the white dwarf effective temperature ($T_{\text{eff}}$), surface gravity ($\log g$), dominant atmospheric element, mass ($M_{\text{WD}}$), parallax, mixing zone (MZ) mass ($M_{\text{MZ}}$), magnetic field ($B_{\text{WD}}$) and their respective errors. Information relevant to the accretion process, such as metal accretion rates, sinking timescales and mass fractions is also provided, as are labels that indicate the presence of dusty or gaseous discs and the presence of binary companions.

To maximise portability, we distribute PEWDD as a \verb|.csv| file and we provide \textsc{python} functions to convert that file into a \textsc{pandas} dataframe, which we recommend as the format for use. The \textsc{python} package\footnote{\href{https://github.com/jamietwilliams/PEWDD}{https://github.com/jamietwilliams/PEWDD}} also contains functions to generate a variety of diagrams exploring the content of PEWDD, examples of which are shown in Sect.\,\ref{sec:plots}. These functions can be called to include only a sub-set of the white dwarfs contained in PEWDD.

A description of each column in PEWDD is provided in Table\,\ref{tab:PWD_format}. We decided to store the information retrieved for a given star from a given publication as a \verb|.json| file, transcribing the published data directly as stated in the literature. These files were converted into a \textsc{pandas} data-frame, with the only manipulation of data being a conversion of all parameters into common units. Individual stars are often referred to by multiple names, and although this means they are easily identifiable in the text of papers, we suggest that \textit{Gaia} DR3 \texttt{source\_id} values are also included when listing white dwarf parameters, as these are unique to each white dwarf. The \verb|Identifier| column is unique to each entry. Some objects do not have a \textit{Gaia} DR3 \texttt{source\_id}, so right ascension and declination coordinates may be required to differentiate white dwarfs. We note that a number of well-studied white dwarfs have multiple published studies included in PEWDD, which may result in multiple abundance measurements for some elements, or abundances for different sub-sets of elements. Hence, investigating all entries for a given white dwarf will provide the most exhaustive overview of the properties of the planetary debris for any individual white dwarf. PEWDD can be cross-matched with other catalogues such as \citet[][hereafter GF21]{GentileFusilloGaia2021}, using the \textit{Gaia} DR3 IDs provided.

For white dwarfs that have multiple sets of abundances in a given paper, for example comparing different modelling assumptions (e.g. \citealt{Xu2017}), we created multiple \verb|.json| files, and include a reference to the specific model in the \texttt{star} column. Published upper limits on abundances are indicated by `--1' in the associated uncertainty, which implies that the inferred accretion rates also must be interpreted as upper limits. The \texttt{comment} column provides notes on the spectra, origin of information, and highlighted mistakes.

In this paper we investigate the observed trends and systematic errors associated with analyses of metal-enriched white dwarfs, and we expect that PEWDD will be a useful tool for the community. It can be used for future comparisons of newly published number abundances, since these can be compared to PEWDD to find similar white dwarfs or accreted bodies to identify the class of material or any outliers. PEWDD will be very useful for detailed investigations into the nature of the accreted material that use number abundances of large samples of metal-enriched white dwarfs \citep[e.g.][]{Bonsor2020,Harrison2021,Buchan2022}. The exclusion of any paper that contains photospheric number abundances is not intentional.

PEWDD will be updated in the future as more abundances are published for metal-enriched white dwarfs. We urge that these abundances are presented in a clear way in a table with associated uncertainties. If accretion rates are provided, sinking timescales should be included. The convective zone (CVZ) mass is an important property that should be quoted if it is used in calculations. We urge that in future abundance studies upper limits for quantities are provided wherever possible to help constrain the nature of the accreted material. The increasing and decreasing accretion rates might be included if sinking timescales are sufficiently long that they are a likely accretion phase, such as in He-dominated atmospheres.

For comparison, the Montreal White Dwarf Database \citep[MWDD,][]{MWDD} contains data and parameters for over 70\,000 white dwarfs of all spectral types, with parameters and spectra publicly accessible. However, MWDD aims to contain information on all white dwarfs, which naturally means it is less complete. Many detailed debris abundance studies focus on a single or a small number of white dwarfs, which are not included within MWDD at the time of writing (e.g. \citealt{KawkaVennes2012, Raddi2015, Farihi2016, Xu2017, GentileFusillo2017, Hollands2021, Izquierdo2021, Klein2021, Johnson2022}). In contrast, PEWDD aims to be complete with respect to studies of planetary enrichment of white dwarfs, which is possible to manage as there are much fewer objects. As such, we believe that PEWDD and MWDD are complementary to each other.

\begin{table*}
    \caption{The format of PEWDD we present in this work.}
    \begin{tabular}{p{0.25\linewidth}|p{0.08\linewidth}|p{0.6\linewidth}}
    \hline
    Column Name & Units & Description \\
    \hline
    \texttt{star} & -- & Name of the white dwarf used in the paper \\
    \texttt{paper} & -- & Paper that the data is taken from \\
    \texttt{identifier} & -- & White dwarf name and paper combined \\
    \texttt{T\_ eff} & K & Effective temperature ($T_{\text{eff}}$) \\
    \texttt{T\_eff\_err} & K & Error on effective temperature \\
    \texttt{logg} & cm\,s$^{-2}$ & Surface gravity ($\log g$) \\
    \texttt{logg\_err} & cm\,s$^{-2}$ & Error on surface gravity. \\
    \texttt{atmosphere} & -- & Dominant atmospheric element, either H or He \\
    \texttt{mass} & M$_{\sun}$ & White dwarf mass ($M_{\text{WD}}$) \\
    \texttt{mass\_err} & M$_{\sun}$ & Error on white dwarf mass \\
    \texttt{RA} & $^{\circ}$ & Right ascension (J2000) \\
    \texttt{Dec} & $^{\circ}$ & Declination (J2000) \\
    \texttt{Gaia\_designation} & -- & \textit{Gaia} Data Release 3 ID (DR3, \citet{Gaia2022}), or if this is not available, \textit{Gaia} DR2 ID instead \\
    \texttt{parallax} & mas & \textit{Gaia} DR3 parallax \\
    \texttt{parallax\_err} & mas & \textit{Gaia} DR3 parallax standard error \\
    \texttt{mixing\_zone\_mass} & g & Mass of the white dwarf mixing zone ($M_{\text{MZ}}$), which is the convective zone or radiative outer layer in cooler and hotter white dwarfs respectively \\
    \texttt{mag\_field} & G & Magnetic field ($B_{\text{WD}}$) \\
    \texttt{mag\_field\_err} & G & Error on magnetic field \\
    \texttt{infrared\_excess} & -- & Presence of infrared excess indicating a dusty disc \\
    \texttt{gas\_component} & -- & Presence of gas emission or circumstellar absorption indicating a disc with a gaseous component \\
    \texttt{binary} & -- & Whether the metal-enriched white dwarf has a binary companion and the \textit{Gaia} DR3 ID of the object. The constraints placed on an object being in a binary system are described in Section \ref{sec:binaries}\\
    \texttt{binary\_separation} & au & The projected separation between the objects \\
    \texttt{origin} & -- & NASA ADS link to paper \\
    \texttt{comment} & -- & Additional comments. Here we include any challenges to the interpretation of photospheric metals being from accreted planetary material \\
    \texttt{log(Z/H(e))} & dex & Photospheric number abundance of element Z relative to dominant atmospheric element, H or He from the `atmosphere' column. See Fig.\,\ref{fig:periodic_table} for details on the detection statistics of individual elements \\
    \texttt{log(Z/H(e))e} & dex & Error on photospheric number abundance. An error of -1 represents an upper limit and an error of 0 represents a poorly constrained abundance \\
    \texttt{acc\_rate\_Z\_steady\_state} & $\mathrm{g\,s^{-1}}$ & The published accretion rate onto the white dwarf surface assuming it is in the steady state phase, where there is an equilibrium between accretion and diffusion \\
    \texttt{acc\_rate\_Z\_increasing} & $\mathrm{g\,s^{-1}}$ & The published accretion rate onto the white dwarf surface assuming it is in the increasing phase, where there is a build-up of material in the white dwarf atmosphere \\
    \texttt{acc\_rate\_Z\_decreasing} & $\mathrm{g\,s^{-1}}$ & The published historical accretion rate onto the white dwarf surface assuming it is in the decreasing phase, where accretion has ceased and material is sinking out of the atmosphere. This is what the accretion rate would have been in the steady state phase \\
    \texttt{time\_since\_acc\_ended} & s & The time since accretion ended that the decreasing phase accretion rate is measured at \\
    \texttt{sinking\_time\_Z} & s & The metal diffusion timescale \\
    \texttt{mass\_fraction\_Z} & -- & The mass fraction of each metal in the accreted material assuming steady state accretion, defined as the metal's accretion rate divided by the total accretion rate \\
    \hline
    \end{tabular}
    \tablefoot{Each row in this table is a column in PEWDD for every included metal-enriched white dwarf. If a paper does not contain the relevant information, the associated column for that object is left blank. An example entry is shown in Table\,\ref{tab:database_example}.}
    \label{tab:PWD_format}
\end{table*}

\begin{table*}
    \caption{An example of a database entry.}
    \centering
    \begin{tabular}{l|l}
    \texttt{Star} & \texttt{SDSS1228+1040}\tablefootmark{a} \\
    \texttt{Paper} & \texttt{Koester 2014} \\
    \texttt{Identifier} & \texttt{SDSS1228+1040 Koester 2014} \\
    \texttt{T\_eff} & \texttt{20713} \\
    \texttt{T\_eff\_err} & \texttt{281} \\
    \texttt{logg} & \texttt{8.15} \\
    \texttt{logg\_err} & \texttt{0.089} \\
    \texttt{atmosphere} & \texttt{H} \\
    \texttt{mass} & \texttt{0.705} \\
    \texttt{mass\_err} & \texttt{0.051} \\
    \texttt{RA} & \texttt{187.249726} \\
    \texttt{Dec} & \texttt{10.675846} \\
    \texttt{Gaia\_designation} & \texttt{Gaia DR3 3904415787947492096} \\
    \texttt{parallax} & \texttt{7.763378} \\
    \texttt{parallax\_err} & \texttt{0.066206} \\
    \texttt{infrared\_excess} & \texttt{disc} \\
    \texttt{gas\_emission} & \texttt{gas} \\
    \texttt{origin} & \texttt{https://ui.adsabs.harvard.edu/abs/2014A\%26A...566A..34K/abstract} \\
    \texttt{[Si/H(e)]} & \texttt{-5.2} \\
    \texttt{[Si/H(e)]e} & \texttt{0.1} \\
    \texttt{[C/H(e)]} & \texttt{-7.6} \\
    \texttt{[C/H(e)]e} & \texttt{0.1} \\
    \texttt{Acc\_rate\_Si\_steady\_state} & \texttt{52480746.024977} \\
    \texttt{Acc\_rate\_C\_steady\_state} & \texttt{169824.365246} \\
    \end{tabular}
    \tablefoot{All empty columns have been removed for clarity.\\
    \tablefoottext{a}{SDSS\,J122859.93+104032.9 (otherwise known as SDSS\,1228+1040; \citealt{Koester2014_DAZ}).}}
    \label{tab:database_example}
\end{table*}

\section{Trends in metal-enriched white dwarf properties}\label{sec:generic}

\subsection{Atmospheric composition}

\begin{figure}
    \centering
    \includegraphics[width=\columnwidth]{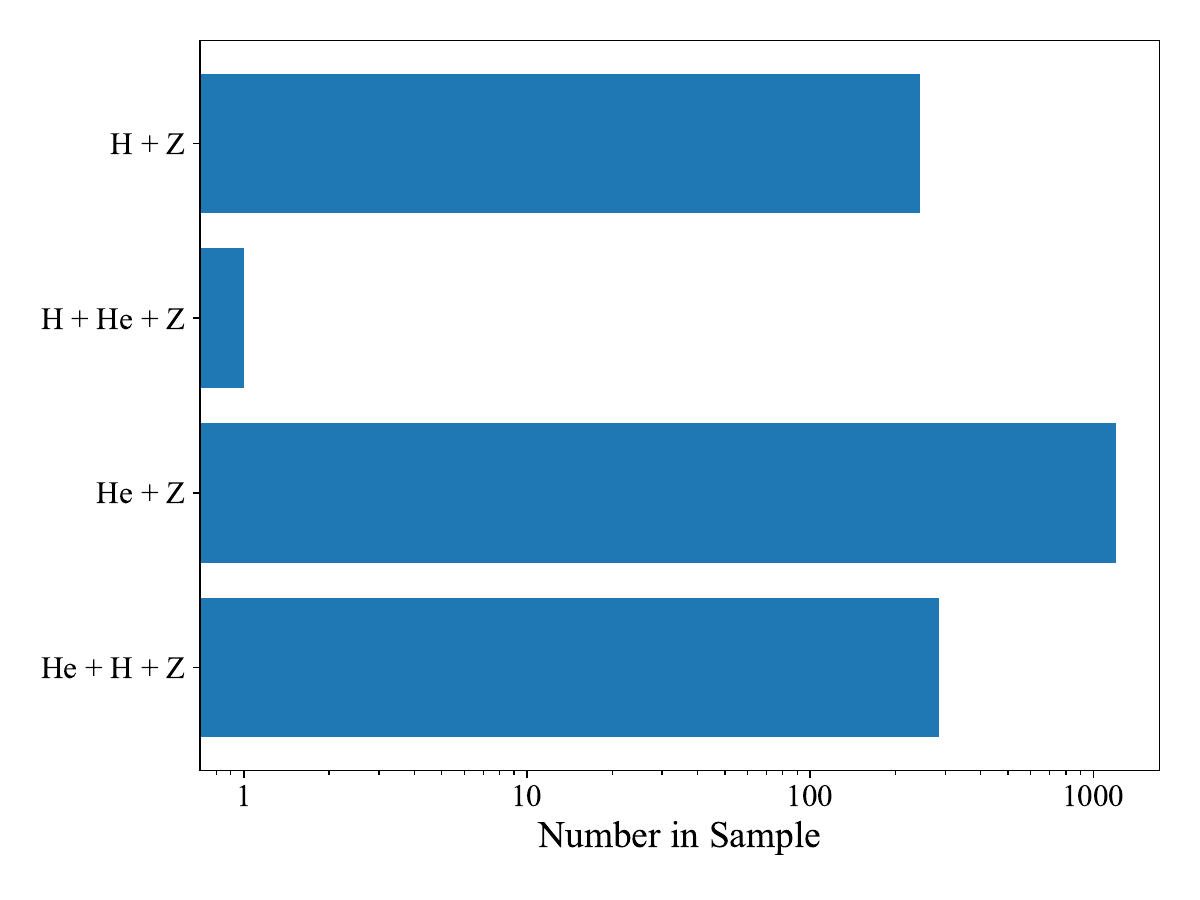}
    \caption{The atmospheric composition of the white dwarfs in PEWDD. “H + Z" and “He + Z" have atmospheres dominated by H and He, respectively, with traces of metals. The white dwarfs with “H + He + Z" and “He + H + Z" have H- or He-dominated atmospheres with traces of both metals, and He or H, respectively.}
    \label{fig:spectral_types}
\end{figure}

Metal-enriched He-dominated white dwarfs outnumber their H-dominated counterparts in PEWDD (Fig.\,\ref{fig:spectral_types}), even though H-atmosphere white dwarfs are much more common than their He-atmosphere counterparts \citep{OBrien2024}. The reduced opacity of He compared to that of H allows for much lower photospheric metal abundances to be detected in He-atmospheres. Conversely, the diffusion timescales on which metals sink out are much longer in He-dominated white dwarfs ($\simeq10^4-10^6$\,yr) than in H-dominated white dwarfs ($\simeq\mathrm{days}$ to years; \citealt{Koester2009_WDs}), implying that metals from past accretion episodes can still be detected in He-dominated white dwarfs, whereas metal-enrichment in H-dominated white dwarfs usually implies ongoing accretion. Both effects are responsible for boosting the number of metal-enriched He-dominated white dwarfs\footnote{In hot ($T_\mathrm{eff}\ga20\,000$\,K) H atmospheres, radiative levitation can balance gravity and hence sustain metals for prolonged periods of time, however, hot and young white dwarfs are relatively rare within the overall white dwarf population. The presence of photospheric metals from accretion and radiative levitation can be distinguished by modelling the predicted abundances from radiative levitation alone \citep{Chayer2014,Koester2014_DAZ}.}. Detections of H in He-dominated atmospheres could be due to accretion of water-rich material \citep[e.g.][]{Wolff2000,GentileFusillo2017} or white dwarf spectral evolution \citep{KoesterKepler2015, Bedard2024}. Vice versa, in rare instances, He may be accreted from a giant planet into the atmosphere of an H-dominated white dwarf.

\subsection{Statistics of metal detection}

\begin{figure*}
    \centering
    \includegraphics[width=\textwidth]{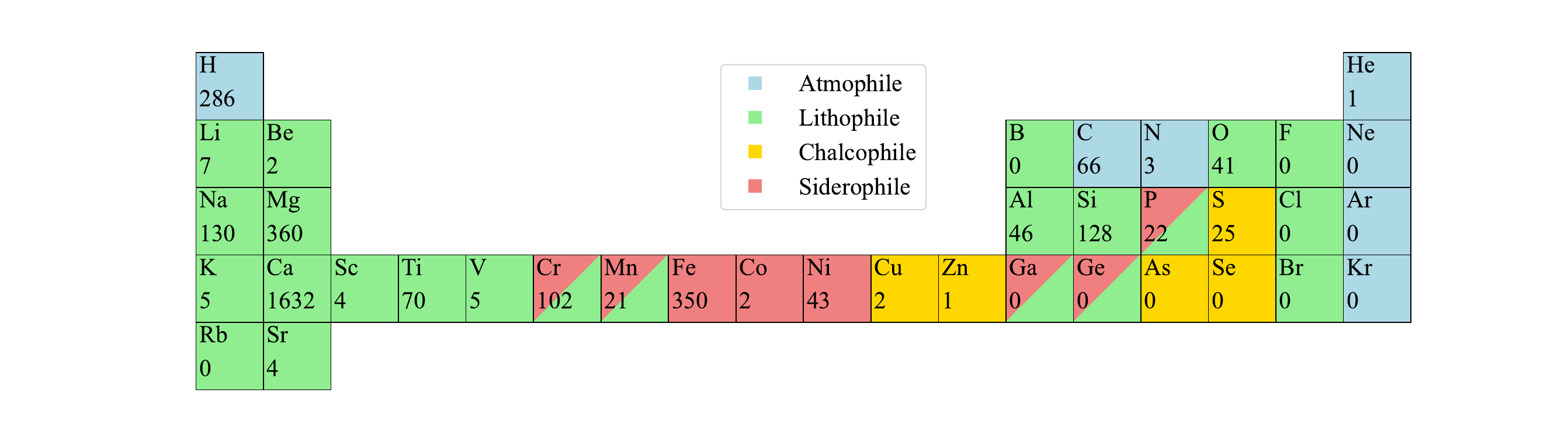}
    \caption{A periodic table highlighting the frequency of metal detection in metal-enriched white dwarfs. The number in each box represents how many white dwarfs have detections of that photospheric metal. The colours represent the Goldschmidt classification \citep{Goldschmidt1937}, which groups metals into their geochemical preferences with atmophiles (gas-loving; blue), lithophiles (silicate-loving; green), chalcophiles (sulfur-loving; yellow) and siderophiles (iron-loving; red). Metals that have two colours within their box exhibit both siderophilic and lithophilic characteristics. The box containing H indicates the number of He-atmosphere white dwarfs with  trace-H detections and vice versa for the box with He.}
    \label{fig:periodic_table}
\end{figure*}

\subsubsection{Geochemical properties}

A total of 24 different metals have been detected in white dwarf photospheres, and the number of published detections per element is illustrated as a periodic table in Fig.\,\ref{fig:periodic_table}. The coloured boxes in Fig.\,\ref{fig:periodic_table} represent the Goldschmidt geochemical classes of metals \citep{Goldschmidt1937}, and can indicate the type of material that has been accreted. Lithophiles (silicate-loving) such as Mg, Si and Ca tend to form silicate rocky material, and can be used to trace exo-planetary mantles and crusts. Siderophiles (iron-loving) like Co or Ni interact with Fe and sink towards the centre of bodies, so can be used to identify core material. Some metals such as P, Cr and Mn can show both lithophilic and siderophilic characteristics since they can exist bonded to both iron and silicates. Chalcophiles (sulfur-loving) react with S and do not sink into the core, so are tracers of mantle material. The relative abundance of these types of metals in accreted material can determine whether the body is core-, mantle- or crust-rich, and hence if the body is differentiated or primitive \citep{Bonsor2020}. Atmophiles (gas-loving) are volatile metals and are often in a gaseous state, leading to depletion of this type in rocky material. This explains the low number of white dwarfs that show atmophile species enrichment. Additionally, weaker absorption features along with interstellar line contamination mean that these detections are often more difficult \citep{Wilson2016}.

\subsubsection{Correlations with effective temperature}

From Fig.\,\ref{fig:periodic_table}, it is apparent that the number of detections varies hugely between the individual elements. The detection of an element depends on its abundance in the white dwarf atmosphere, and the strength of its spectral features in the observed wavelength range. In turn, the latter strongly depends on the ionisation and excitation balance within the atmosphere, and hence on the temperature of the star. We note that, with a few exceptions \citep[e.g.][]{Xu2017,Gaensicke2019} bulk abundances across various metal enriched white dwarfs do not vary hugely, analogous to the small spread of stellar abundances \citep{Payne1925}. Instead, the differences in the detectability of a given metal reflect the conditions in the white dwarf atmosphere. 

\begin{figure*}
    \centering
    \includegraphics[width=\textwidth]{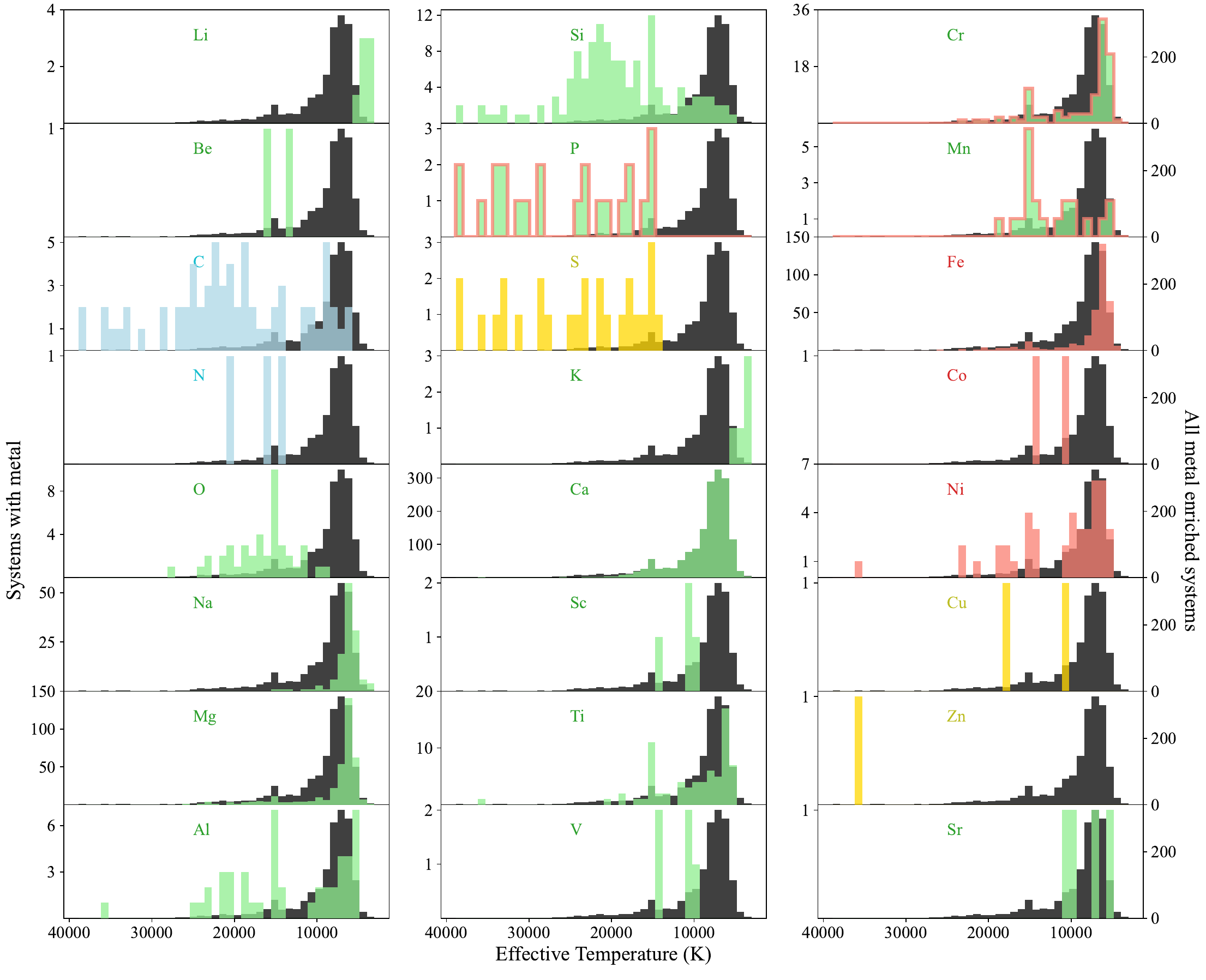}
    \caption{The number of detections of individual elements as a function of $T_{\text{eff}}$, with colours related to the Goldschmidt classification, as described in Fig.\,\ref{fig:periodic_table}. The black histograms show the temperature distribution for all metal detections, normalised to match that of the individual elements.}
    \label{fig:T_effhist}
\end{figure*}

The distribution of detections of the individual elements is shown as a function of $T_\mathrm{eff}$ in Fig.\,\ref{fig:T_effhist}. As the \ion{Ca}{II} H/K resonance lines are the strongest features in the optical spectra of white dwarfs with $T_\mathrm{eff}\lesssim18\,000$K, Ca is the most common element found in PEWDD. It is detected in 94.5~per cent of all metal-enriched white dwarfs~--~even though the mass fraction of Ca is only $\simeq1.7$~per cent in the bulk Earth \citep{McDonough2000}. Therefore, with the exception of the hottest white dwarfs, the distribution of Ca detections as a function of white dwarf temperature is nearly identical to that of all metals. The next most frequently detected metals are Mg ($\simeq15.5$~per cent of the bulk Earth) and Fe ($\simeq31.9$~per cent of the bulk Earth), both of which have strong lines at optical wavelengths. Other common elements such as O ($\simeq29.7$~per cent of the Bulk Earth) and Si ($\simeq16.1$~per cent of the bulk Earth) have no or weak lines in the optical spectra of cool white dwarfs ($T_\mathrm{eff}\lesssim13\,000$\,K), and are therefore only rarely detected, but are more commonly found in UV observations of hotter white dwarfs. The same is true for volatile elements (C, N, S, P). For hotter white dwarfs, the far-UV is generally much better suited than the optical, as it provides a plethora of strong transitions of most elements. However, the \textit{Hubble Space Telescope} (\textit{HST}) is currently the only facility providing access to far-UV spectroscopy, hence limiting the number of white dwarfs observed in that wavelength range. The detectability of metals is strongly correlated to $T_{\text{eff}}$, as the temperature at which the resonance transitions can be ionised will give the highest chance of that metal being observed.

\begin{figure*}
    \centering
    \includegraphics[width=\textwidth]{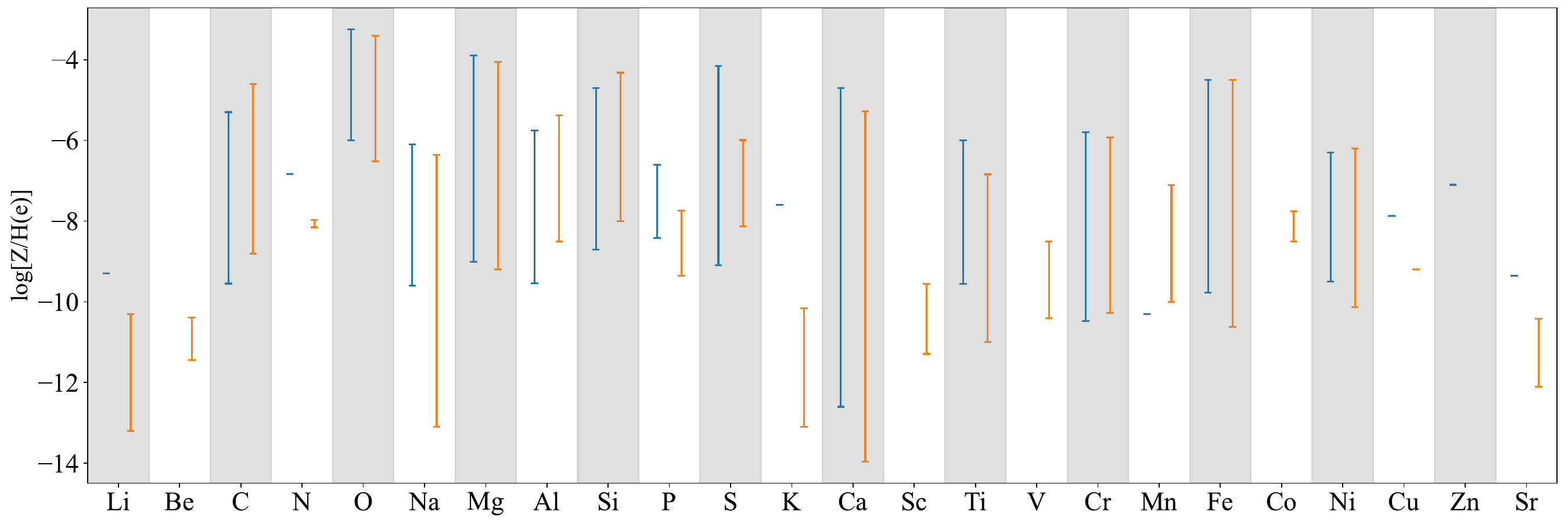}
    \caption{Upper and lower abundance bounds for all metals detected in H- (blue) and He- (orange) dominated white dwarf atmospheres. The absence of a bound means that the metal has not been observed (Be and Co in a H-dominated white dwarf atmosphere). The single dashes represent metals that have only been detected in a single white dwarf of a certain atmosphere composition.}
    \label{fig:minmaxelements}
\end{figure*}

The upper and lower bounds on the measured photospheric abundances of the various metals detected in white dwarfs are shown in Fig.\,\ref{fig:minmaxelements}. The volatiles N, P and S have high lower bounds, suggesting that they require a substantial photospheric abundance to be observed. In contrast, Li, Be and K are detected only in a small number of stars, and always at low abundances, implying that their mass fractions in the accreted material are small. Some metals vary over many orders of magnitude, implying that these can be detected over almost all accretion rates. The metal with the largest detected abundance is O in WD\,1536+520 at $\log(\text{O}/\text{He})= -3.40 \pm 0.20$, which is likely accreting a water-rich body \citep{Farihi2016}. The ultra-cool WD\,J2147$-$4035 holds the record of the lowest observed number abundance in PEWDD, with $\log(\text{Ca}/\text{He}) = -13.96 \pm 0.2$ \citep{Elms2022}, i.e. metal abundances within PEWDD span over ten orders of magnitude.

\begin{figure}
    \centering
    \includegraphics[width=\columnwidth]{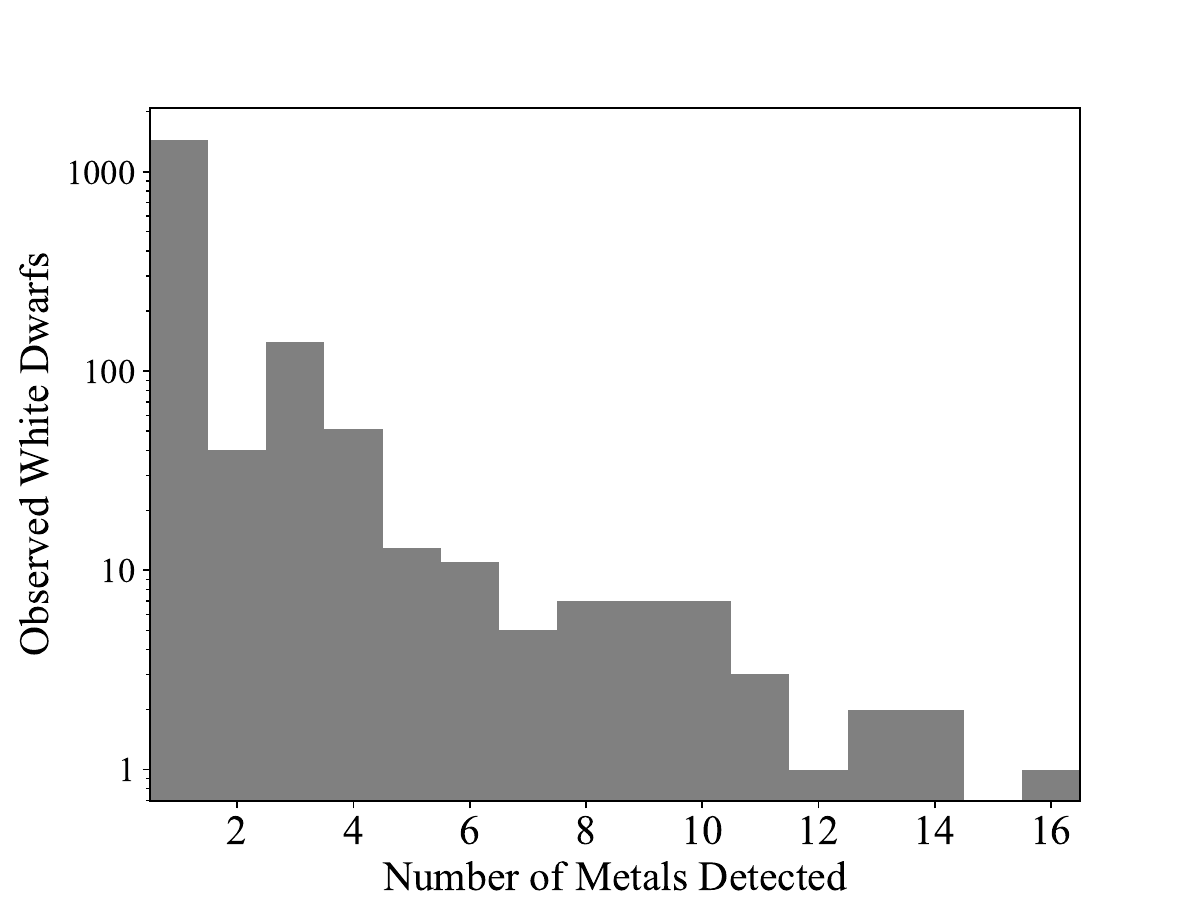}
    \caption{The frequency of metal-enriched white dwarfs as a function of detected photospheric metals. White dwarfs with multiple abundance analyses are combined to find the total number of metals for each object. H and He are not counted.}
    \label{fig:hist}
\end{figure}

The number of metals observed in a white dwarf atmosphere is often limited by the resolution of the spectra. Due to the strength of the \ion{Ca}{II} line as a powerful tracer of metal-enrichment, even low resolution spectra can determine whether a white dwarf is accreting. 
The larger number of low-resolution spectra contributes to fewer metals being observed in many white dwarfs, often only Ca\footnote{Additionally, many large scale surveys focus exclusively on the detection of Ca (e.g. \citealt{Kepler2015,Kepler2016,Kepler2021}; \citealt{Coutu2019}; \citealt{Kilic2020}) and overlook absorption features that indicate the presence of other metals.}. These factors largely explain the strong bias towards the large number of only Ca detections evidenced in Figs.\,\ref{fig:periodic_table} and \ref{fig:hist}. Most of these white dwarfs are very likely accreting material containing dozens of further metals.

Only 25\,per cent of the stars within PEWWD have more than one element detected, among which we note a  deficiency in metal-enriched white dwarfs with only two detected photospheric metals (Fig.\,\ref{fig:hist}). This is a consequence of better quality data revealing usually several additional species beyond Ca, particularly Mg and Fe (e.g. \citealt{Hollands2017}; \citealt{Blouin2020}). For white dwarfs with four or five unique species,  Na, Si and Cr become more common. Larger numbers of elements are found in white dwarfs that are strongly enriched \citep[e.g.][]{Raddi2015,Farihi2016,GentileFusillo2017}, have high resolution spectra \citep[e.g.][]{Klein2010,Zuckerman2011} or cool He-dominated stars with moderately large amount of metals detectable at low resolution \citep[e.g.][]{Hollands2017}. 

As shown above, metal-enriched white dwarfs exhibit a large variety with respect to how many, and which elements can be detected, and their individual abundances. The most diversely metal-enriched white dwarf is GD\,362 with 16 metals in its atmosphere (\citealt{Zuckerman2007}; \citealt{Xu2013}), with the accreted material resembling bulk Earth in terms of its composition. Photospheric Zn is only detected in the hot white dwarf HS\,0209+0832 \citep[][Williams et al. in prep]{Wolff2000}. Be is the most recently discovered photospheric metal \citep{Klein2021}. The next metals that may be detected considering the strength of their absorption features and bulk Earth abundance are B, Ga and Ge.

\subsubsection{Metals in hot white dwarfs}

Our knowledge regarding planetary enrichment in hot ($T_\text{eff}\gtrsim 25\,000$\,K) and young white dwarfs is so far limited. Most hot white dwarf exhibit photospheric metals \citep{Barstow2003} including exotic elements such as Xe and Kr \citep{Werner2012}, Br and Sb \citep{Werner2018} and Cs \citep{Chayer2023}. However, the origin of these elements remains debated: historically, radiative levitation counteracting the downward diffusion of metals \citep[e.g.][]{Chayer1995} has been the favourite model to explain the detection of photospheric metals in hot white dwarfs, even though it was recognised early on that the predicted abundances of individual elements did not agree with the observations \citep{Chayer1994,Wolff2000}. At very high temperatures ($T_\text{eff}\gtrsim75\,000$\,K) gravitational settling is very likely incomplete \citep{Werner2015, Werner2018}, preventing the study of planetary enrichment in these youngest ($\lesssim5$\,Myr) white dwarfs. For intermediate temperatures, $T_\text{eff}\simeq25\,000-75\,000$\,K), \citet{Barstow2014} demonstrated that external enrichment plays an important role for the photospheric abundances of white dwarfs, however, the effect of radiative levitation has to be taken into account when establishing the abundances of accreted material \citep{ChayerDupuis2010, Koester2014_DAZ}.

In addition to the physical complications within the atmospheres of hot white dwarfs, the origin of material enriching them (where an unambiguous case can be made) is not well established. The distance from the white dwarf at which dust particles will sublimate increases with $T_\mathrm{eff}$. Depending on the sublimation temperature of the dust \citep{Rafikov2012}, the location of the sublimation radius moves beyond the tidal disruption radius for $T_\mathrm{eff}\simeq25\,000-30,000$\,K \citep{vonHippel2007}, and it is thought that hotter stars cannot host a dusty disc that could feed material onto the white dwarf over extended periods of time. However, hot white dwarfs may photo-evaporate close-in giant planets \citep{Schreiber2019, Gallo2024}. WD\,J091405.30+191412.25 \citep{Gaensicke2019} is currently the only confirmed system, but
several other hot stars show unusual photospheric enrichment (e.g. GD394, \citealt{Wilson2019}; HS\,0209+0832, \citealt{Wolff2000}).

\section{Relative abundances} \label{sec:plots}

Discussing the nature of the accreted material often involves inspecting diagrams involving two different abundance ratios, comparing the measurements obtained from modelling the spectroscopy of white dwarfs to those of a reference composition (e.g. Fig.\,7 in \citealt{Gaensicke2012}; Fig.\,3 in \citealt{Swan2019}; Fig.\,9 in \citealt{Izquierdo2021}; Fig.\,2 in \citealt{Hollands2021}; Fig.\,11 in \citealt{Klein2021}). In this section, we refer to ``abundance'' as the parent body composition, which is calculated by correcting the photospheric abundances by the relative diffusion timescales. These diagrams often, but not always involve the same denominator of the ratios on both axes. Using meteorite abundance data (bulk Earth and bulk Silicate Earth, \citealt{McDonough2000}; solar abundances and CI chondrites, \citealt{Lodders2003} and various meteorite classes, \citealt{Nittler2004}) as abundance references, the type of the accreted planetary material can hence be constrained. With measured accretion rates, Eq.\,\ref{eq:number_abundance} gives an example of calculating the abundance ratios with respect to Si, where $A$ is the atomic mass, though other reference metals such as Mg or Fe are also frequently used. We provide code and a guide to use the functions in order to quickly construct this type of diagram with a set of white dwarfs and reference compositions\footnote{When comparing to Solar System objects, traditionally Mg, Si or Fe are used, but any metal can be selected as a reference} chosen by the user \citep{Allegre2001,McDonough2000,Lodders2003,Nittler2004,Asplund2005}.

\begin{equation}
    \log\left(\frac{N\text{(Z)}}{N\text{(Si)}}\right) = \log\left(\frac{\dot{M}\text{(Z)}}{\dot{M}\text{(Si)}} \frac{A\text{(Si)}}{A\text{(Z)}} \right)
    \label{eq:number_abundance}
\end{equation}

\begin{figure}
    \centering
    \includegraphics[width=\columnwidth]{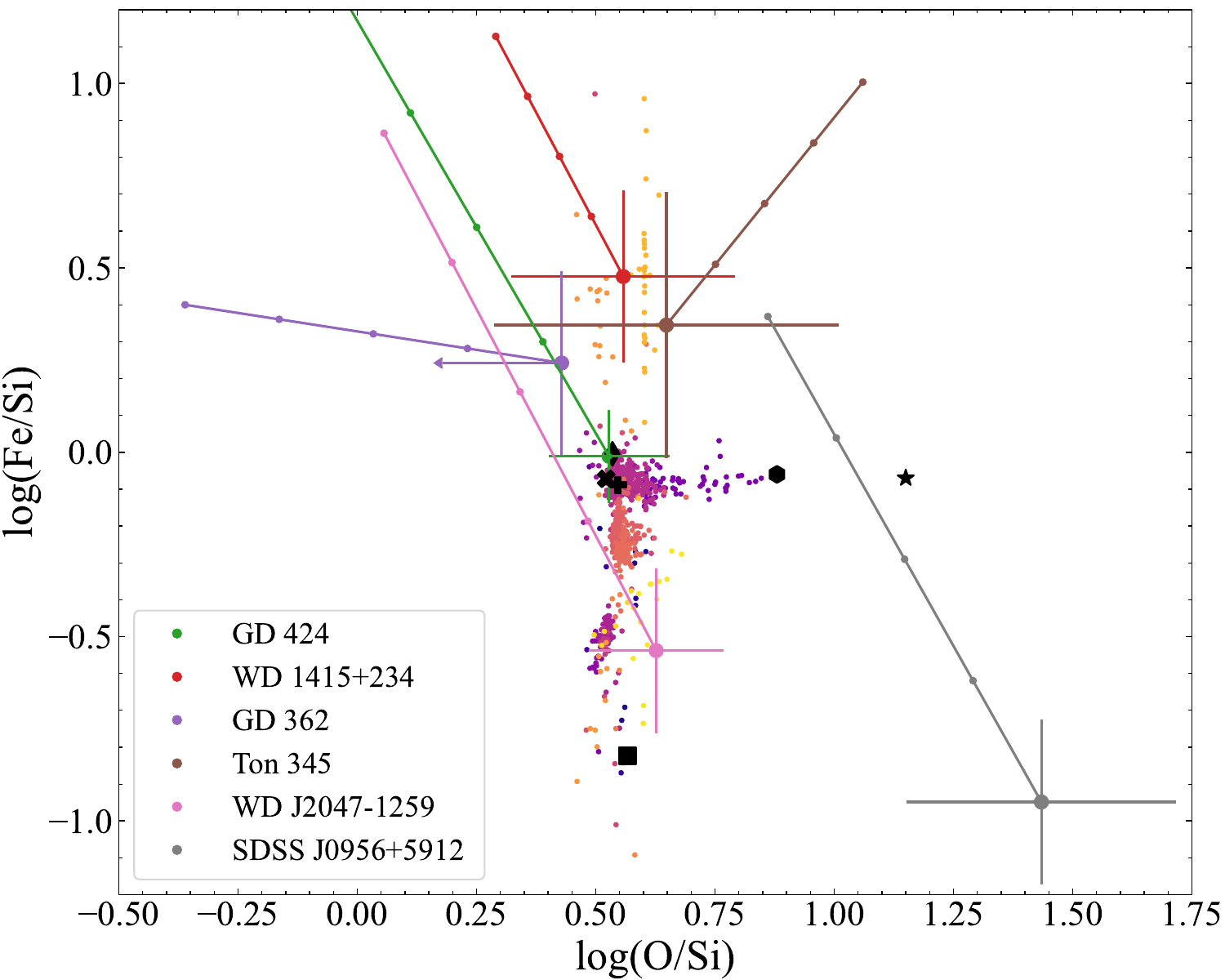}
    \caption{An abundance ratio diagram comparing O and Fe with the reference metal Si. The coloured dots are different classes of meteorites \citep{Nittler2004} and the black symbols are large Solar System objects (\citealt{McDonough2000}; \citealt{Lodders2003}, CI\,=\,chondrites and BSE\,=\,bulk silicate Earth). Six metal-enriched white dwarfs with diverse abundances are shown as examples (GD\,424, \citealt{Izquierdo2021}; WD\,1415+234, \citealt{Doyle2023}; GD\,362, \citealt{Xu2013}; Ton\,345, \citealt{Wilson2015}; EC\,20444–1310 also known as WD\,2047$-$1259, \citealt{Hoskin2020}, and SDSS\,J095645.14+591240.6, \citealt{Hollands2022}). The diagonal lines track the abundances back through time, assuming the star is observed in the decreasing phase. Each small point on the diagonal line represents the corrected abundance of the accreted body if accretion had ceased one Si sinking timescale ago. Diagrams like this can be created with any combination of metals available within PEWDD and the \textsc{Python} code that we provide. The evolutionary lines require that the sinking timescales were published as well.}
    \label{fig:number_abundance}
\end{figure}

\noindent
An example of an abundance ratio diagram comparing O/Si to Fe/Si is shown in Fig.\,\ref{fig:number_abundance}.  Comparing these two ratios is useful since all three are major rock-forming elements and Fe may also occur in metallic form in the differentiated cores of larger planetary bodies. 

It is evident from inspection of Fig.\,\ref{fig:periodic_table} that although O is a major building block of terrestrial planets, it is difficult to detect in white dwarf atmospheres. GD\,362 only has an upper limit on O which is represented by an arrow pointing along a O-poor direction. SDSS\,J095645.14+591240.6 is located far from the region occupied by Solar System objects, since it is observed in the decreasing phase \citep{Swan2023}. Even though the uncertainties on the abundance ratios measured from the analysis of white dwarfs are much larger than those of the solar-system reference compositions,  these types of diagrams are useful to infer the types of parent bodies accreted by white dwarfs, and the work carried out so far clearly demonstrates that many white dwarfs disrupted planetary objects with compositions similar to what is common within the inner solar system.

For white dwarfs with He-dominated atmospheres, the phase of accretion must be considered. In a simplistic model assuming that an accretion event starts instantaneously, provides a constant accretion rate for a specific amount of time, and then switches off instantaneously, there are three well-defined phases \citep{Koester2009_WDs}: the increasing (build-up) phase, steady-state phase and the decreasing phase (a brief description is provided in Table\,\ref{tab:PWD_format} and a more detailed discussion in \citealt{Farihi2016rev}). In the decreasing phase, the abundance ratio measured at a time $t$ is related to that at time time that accretion stopped, $t_0$ via

\begin{equation}
    \log \left(\frac{N(\text{Z},t)}{N(\text{Si},t)}\right) = \log \left(\frac{N(\text{Z},t_0)e^{t/\tau_{\text{Z}}}}{N(\text{Z},t_) e^{t/\tau_{\text{Si}}} }\right)
    \label{eq:sinking_timescales}
\end{equation}

\noindent
Metals with short (long) sinking timescales would have had larger (smaller) abundances in the accreted material compared to the measured photospheric abundances, since heavier metals would have sunk out of the bottom of the convection zone faster (slower).  If the sinking time scales of a white dwarf are comparable to the estimated disc life times ($\simeq10^4-10^6$\,yr, \citealt{Girven2012, Cunningham2021}), the possibility of the system being in one of these three phases should be explored, as illustrated by the tracks in Fig.\,\ref{fig:number_abundance}. Each small point on these tracks represents the abundances of the parent body if accretion had ceased one Si sinking timescale ago, using the time-dependent corrections from Eq.\,\ref{eq:sinking_timescales}. In the case of WD\,J2047$-$1259, where $\tau_{\text{Fe}}=2.6\times10^3\,$yrs and $\tau_{\text{Si}}=4.7\times10^3\,$yrs, more Fe than Si has sunk out of the atmosphere in the time elapsed since accretion turned off, so tracking the abundances back in time leads to larger Fe abundance in the parent body than the photospheric measurements suggest. The opposite is seen for $\log(\mathrm{O/Si})$, where $\tau_{\text{O}}=7.0\times10^3\,$yrs: less O has sunk than Si, so the accreted body  appears more O-rich in the present-day measurement than it actually is \citep{Hoskin2020}. The length and direction of these tracks depends on the ratio of the sinking times, e.g. the track of Ton\,345 is nearly perpendicular to that of WD\,J2047$-$1259 since its sinking timescales are $\tau_{\text{O}}=9.7\times10^3\,$yrs, $\tau_{\text{Si}}=12\times10^3\,$yrs and $\tau_{\text{Fe}}=8.7\times10^3\,$yrs, meaning that the body will be more O-rich than the photospheric measurement if accretion has stopped some time ago. Before claiming that a white dwarf accretes a planetesimal with an exotic composition, it is advisable to investigate if the system could be in the decreasing phase, and whether accretion ending a few diffusion time scales ago would bring it closer to counterparts in the Solar System~--~to date, the majority of white dwarfs with detailed analyses fall within these reference compositions (\citealt{JuraYoung2014}; \citealt{Trierweiler2023}). Whereas the inspection of  abundance ratio diagrams is a visual and intuitive diagnostic, more sophisticated methods employ Bayesian models to infer both the most likely accretion phase and abundances (e.g. \citealt{Harrison2021}, \citealt{Swan2023}).

\begin{figure*}
    \includegraphics[width=0.49\textwidth]{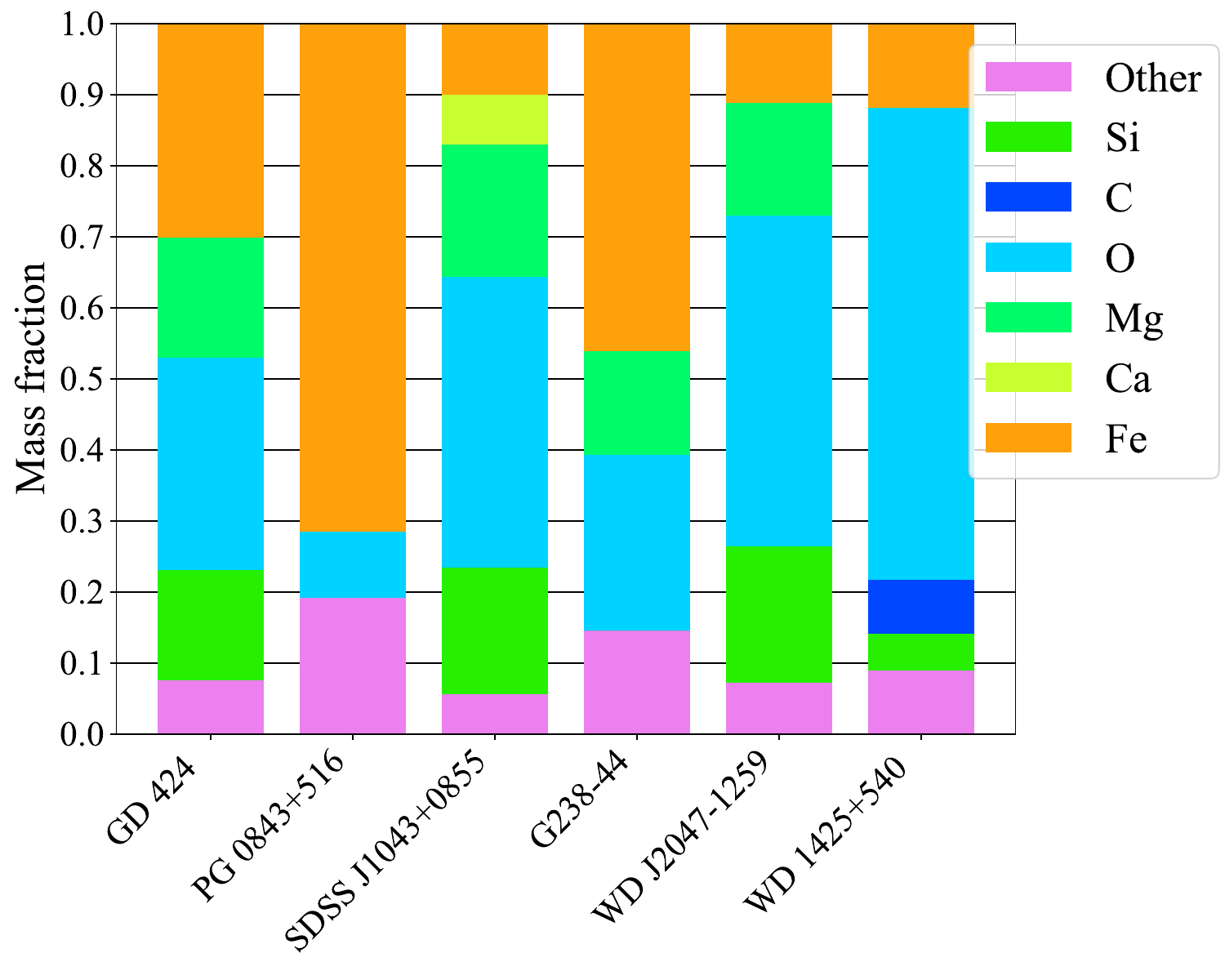}
    \includegraphics[width=0.49\textwidth]{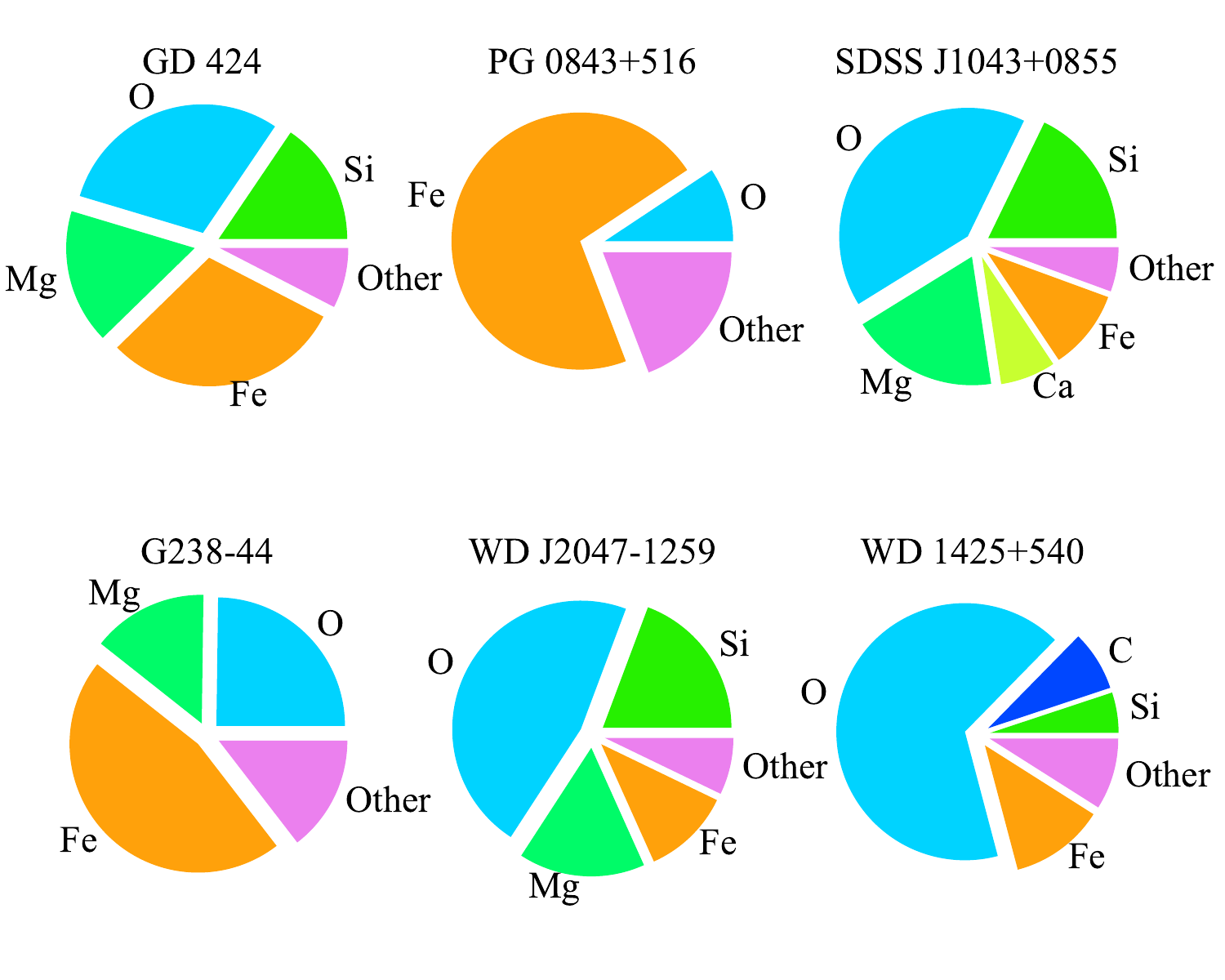}
    \caption{Mass fractions for six parent bodies accreted onto white dwarfs, shown as bar charts on the left and pie charts on the right, illustrate the diversity in compositions: bulk Earth for GD\,424 \citep{Izquierdo2021}; core-rich material for PG\,0843+516 \citep{Gaensicke2012}; mantle- and crust-rich material for SDSS\,J104341.54+085558.1 \citep{MelisDufour2017}; an unusual composition for G\,238-44, likely a combination of an iron-rich and icy object \citep{Johnson2022}; a hydrated carbonaceous chondrite for EC\,20444–1310 (also known as WD\,J2047--1259; \citealt{Hoskin2020}) and a comet for WD\,1425+540 \citep{Xu2017}. These diagrams were created using PEWDD and \textsc{Python} code that we provide.}
        \label{fig:mass_fractions}
\end{figure*}

If the abundances of the most common elements that make up rocky material have been measured\footnote{If the composition is not consistent with a dry, rocky body additional metals may be required, such as C for an ice-dominated body.} (e.g. O, Si, Fe, Mg), the total accretion rate can be computed, as well as the mass fractions of the individual elements:

\begin{equation}
    X(\text{Z}) = \frac{\dot{M}(Z)}{\Sigma\dot{M}(Z)}
\end{equation}

\noindent
These mass fractions can be instructive if displayed either as bar or pie charts, and Fig.\,\ref{fig:mass_fractions} demonstrates that planetary bodies accreted by white dwarfs represent, just as meteorites in the solar system, a large diversity in their compositions. GD\,424 (assuming it is in the steady-state) is likely accreting material consistent with bulk Earth composition \citep{Izquierdo2021}. The material enriching PG\,0843+516 (in the steady-state phase since sinking timescales in H-dominated atmospheres such as this white dwarf are so low) is enhanced in Fe compared to the bulk Earth material and has likely undergone partial melting and differentiation, with a core fragment being now accreted by the white dwarf \citep{Gaensicke2012}. In contrast, SDSS\,J104341.54+085558.1 is depleted in Fe with respect to the bulk Earth, and hence accreting a mantle fragment \citep{MelisDufour2017}. G\,238-44 has one of the most unusual compositions, rich in O, Mg and Fe, speculated to be the simultaneous accretion of an iron-rich Mercury-like object and an icy Kuiper Belt-like object \citep{Johnson2022}. WD\,J2047$-$1259 (also known as EC\,20444$-$1310; \citealt{Hoskin2020}) has likely accreted a carbonaceous chondrite, particularly rich in O, suggesting the presence of water, but depleted in Fe and Ni, so likely an object that has not undergone differentiation. Finally, WD\,1425+540 \citep{Xu2017} is one of the very few white dwarfs displaying large abundances of volatiles such as C, N, O and observed to have a N abundance similar to that of comets. These mass fraction diagrams can be produced using the code we provide.

\section{Variations in the photospheric abundances} \label{sec:errors}

The photospheric abundances are determined by fitting a synthetic spectrum generated by an atmospheric model to an observed spectrum. Applying this method to different data sets of the same star, or using different models may lead to different results, for a number of underlying reasons: (1) fitting spectroscopy or photometry plus astrometry provides the fundamental atmospheric parameters of a white dwarf, $T_{\text{eff}}$ and $\log g$, and changes in these parameters directly affect the resulting abundances \citep{Izquierdo2023}. However, as long as the changes in $T_\text{eff}$ and $\log g$ are sufficiently small, the relative abundances, i.e. element ratios, remain largely unaltered. (2) Different input physics within the model atmosphere code can affect both $T_\text{eff}$ and $\log g$, as well as the derived abundances \citep[e.g.][]{koester2010,Blouin2020}. Great care has to be taken specifically in the case of heavily enriched white dwarfs to account appropriately for the blanketing effect of spectral lines across the entire wavelength range of the model. (3) The observations used to measure the abundances that are included in PEWDD cover different wavelength ranges, spectral resolutions, instruments and S/Ns. As such, PEWDD is a heterogeneous database which provides the opportunity to investigate the impact that different models and data sets have on the photospheric abundances.

Another important aspect where authors use different approaches is the estimation of the uncertainties of the measured abundances. An ad-hoc method is to vary the fit until it is clearly incompatible with the data which typically results in uncertainties of $\simeq0.1-0.3$\,dex \citep[e.g.][]{Gaensicke2012,Koester2014_DAZ}. Alternatively, the differences in the abundances arising from fitting to different spectral lines \citep[e.g.][]{Dufour2010,Xu2019} can be investigated, or statistical errors can be determined \citep[e.g.][]{Tremblay2020}, typically leading to uncertainties of $\simeq0.01-0.05$\,dex. Finally, some authors will attempt to propagate the uncertainties in $T_\text{eff}$ and $\log g$ into the abundance errors. There is, as of now, no agreement on the ``best practice'' to determine abundance errors.

\begin{figure}
    \centering
    \includegraphics[width=\columnwidth]{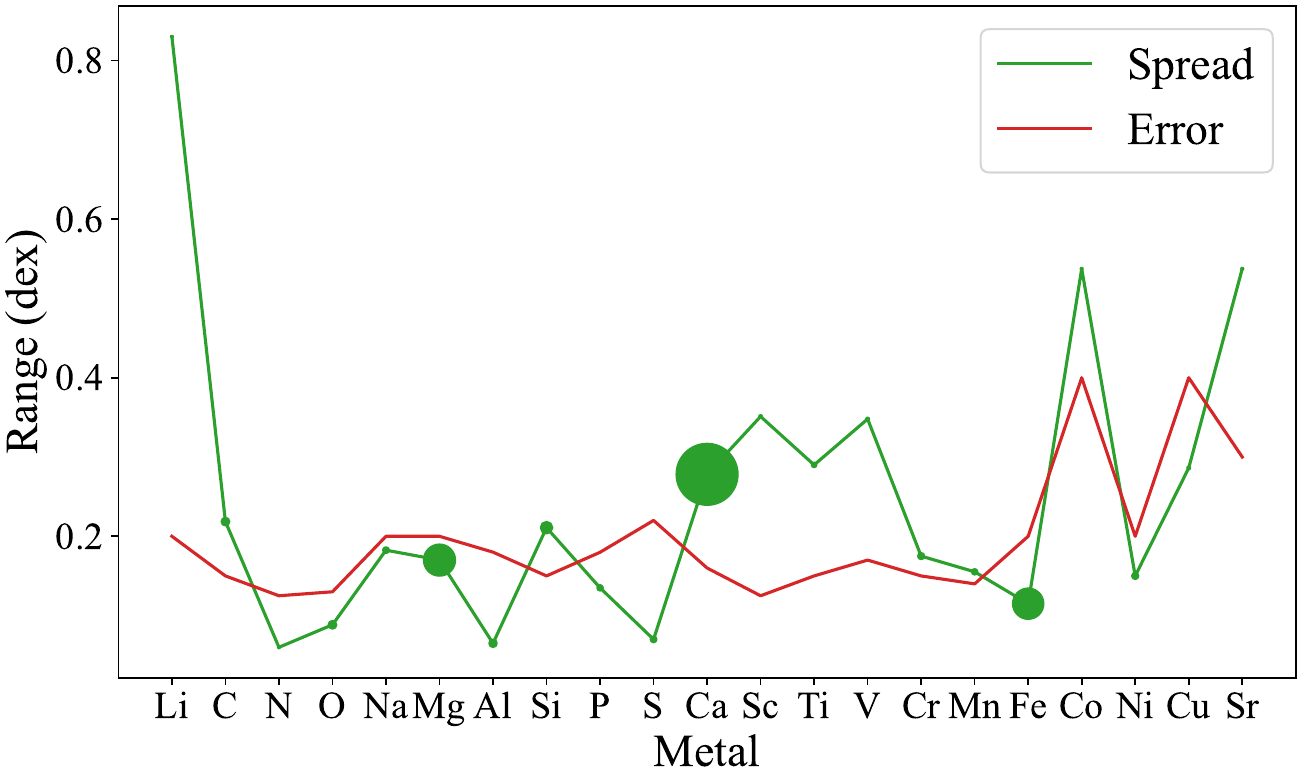}
    \caption{The spread of abundances for each metal detected in a white dwarf atmosphere (green), compared to the median of the quoted errors (red). The size of each point represents the number of measurements that were used to calculate each spread on the metal, with larger points representing more measurements.}
    \label{fig:errors_comparison}
\end{figure}

Here, we used PEWDD to investigate both the variations in  abundances as well as the uncertainties provided in the literature. For that purpose, we define the spread of the abundances as follows: For a white dwarf with multiple measurements within PEWDD, we calculate the standard deviation for all abundances for each metal. Then we compute, for a given element, the median of these standard deviations across all white dwarfs for which that element has multiple measurements to find the spread. We also compute, for each element, the median of all published uncertainties. Figure\,\ref{fig:errors_comparison} compares the spread of the abundances with the median errors. For most metals, the spread varies between 0.1--0.3\,dex, generally consistent with the quoted ad-hoc errors, but substantially larger than the statistically determined errors. Some metals (e.g. Li, Co, Sr) have a very large spread, which is a reflection of the small number of white dwarfs where these elements have been detected, and/or that these metals have comparatively weak spectral lines. The three most frequently detected metals are Mg, Ca and Fe, all three have relatively low spreads, $<0.3$\,dex, similar to the quoted errors.

\begin{figure*}
    \centering
    \includegraphics[width=\textwidth]{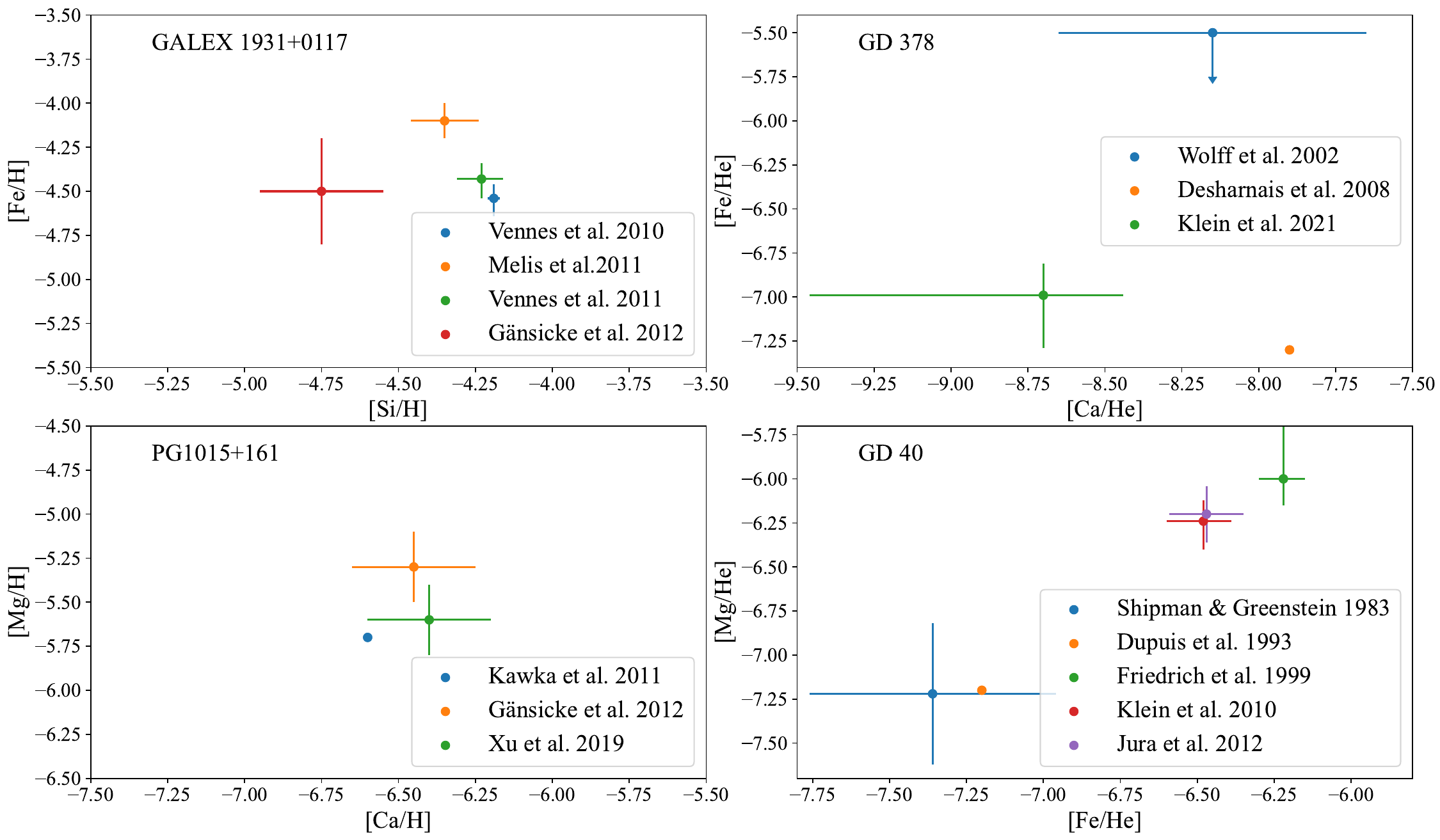}
    \caption{A comparison of abundances taken from different studies of the same white dwarfs, illustrating the range of measurements obtained due to using different observations and models. Each axis is 2\,dex wide. The upper left panel shows Si and Fe photospheric abundances for GALEX\,J193156.8+011745 (\citealt{Vennes2010}; \citealt{Melis2011}; \citealt{Vennes2011}; \citealt{Gaensicke2012}). The upper right panel shows Ca and Fe photospheric abundances for GD\,378 (\citealt{Dupuis1993}; \citealt{Wolff2002}; \citealt{Desharnais2008}; \citealt{Klein2021}). The bottom left panel presents photospheric Ca and Mg abundances for PG\,1015+161 (\citealt{Kawka2011}; \citealt{Gaensicke2012}; \citealt{Xu2019}). The bottom right panel contains photospheric abundances of Fe and Mg for GD\,40 (\citealt{ShipmanGreenstein1983}; \citealt{Dupuis1993}; \citealt{Friedrich1999}; \citealt{Klein2010}; \citealt{Jura2012}).}
    \label{fig:systematics}
\end{figure*}

We pick four well-studied metal-enriched white dwarfs  to illustrate the relatively minor spread of the abundances measured by up to five different works. The sets of Si and Fe abundances measured for the DAZ white dwarf GALEX\,J193156.8+011745 are in fairly good agreement (Fig.\,\ref{fig:systematics}, upper left). The DBAZ white dwarf GD\,378 (Fig \ref{fig:systematics}, upper right) illustrates the improvement in the determination of abundances, where \citet{Desharnais2008} and \citet{Klein2021} detect Fe whereas the data of the earlier study \citet{Wolff2002} was not of sufficient quality. The abundances for Ca vary by a large amount, although they also have large associated errors. Advancements in atmospheric modelling may also contribute to these differences. PG\,1015+161 is an example where the published  abundances agree well within quoted errors (Fig.\,\ref{fig:systematics}, bottom left). Finally, the DBZ white dwarf GD\,40 (Fig.\,\ref{fig:systematics}, bottom right) was subject to analyses over multiple decades, with improvements in observations and modelling leading to abundances converging to an agreed value. 

\section{Accretion rates} \label{sec:accretion}

\subsection{Total accretion rate dichotomy}

Many white dwarfs accrete material with approximately bulk Earth composition \citep{Trierweiler2023} and almost every metal-enriched white dwarf has a detection of Ca (see Fig.\,\ref{fig:periodic_table}). Often the total accretion rate is calculated by scaling from the Ca accretion rate, considering the mass fraction of Ca in bulk Earth of 1.71\,per cent \citep{McDonough2000}. Using published total accretion rates or extrapolating from bulk Earth mass fractions of Ca or Si (16.1\,per cent), we investigate the relationship between $T_{\text{eff}}$ and total accretion rate, similar to \citet{Blouin2022}.

\begin{figure*}
    \centering
    \includegraphics[width=\textwidth]{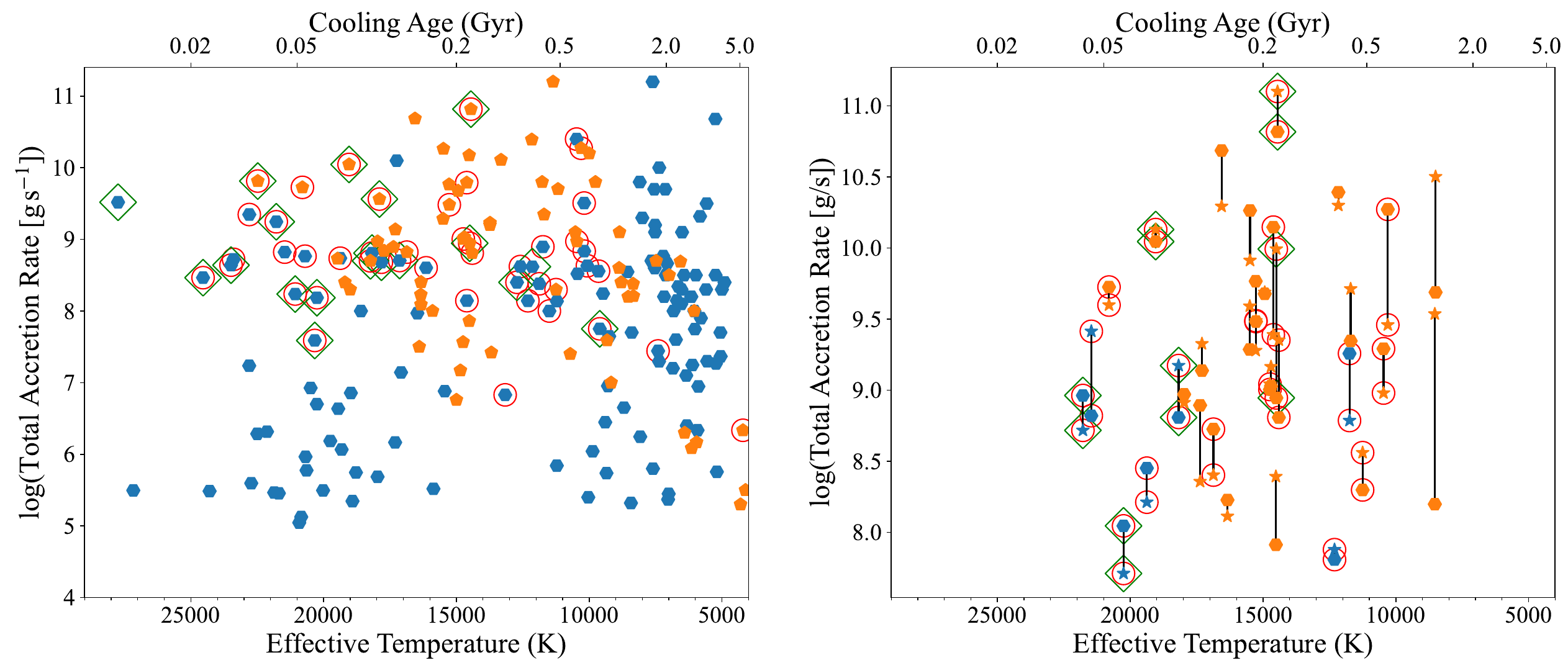}
    \caption{Total accretion rates onto white dwarfs with H- (blue hexagons) and He- (orange pentagons) dominated atmospheres as a function of temperature. The left panel shows total accretion rates extrapolated from that of $\dot M_\mathrm{Ca}$ (or $\dot M_\mathrm{Si}$ if Ca was not detected), assuming that the accreted material has bulk Earth abundance \citep{McDonough2000} with a Ca (Si) mass fraction of 1.71\,per cent (16\,per cent). The red rings and green diamonds highlight the systems where infrared excesses and gas emission have been detected respectively, implying the presence of a circumstellar disc. The right panel compares the extrapolated total accretion rate to the sum of the accretion rates of O, Mg, Si and Fe, which make up approximately 95\,per cent of the bulk Earth, represented by stars.}
    \label{fig:T_eff_vs_acc_rate}
\end{figure*}

To assess the validity of total accretion rates computed from a single detected element, we compare in the right panel of Fig.\,\ref{fig:T_eff_vs_acc_rate} two methods of calculating the total accretion rate. The first, represented by the hexagons is scaling the accretion rate using Ca and the second is summing up the accretion rates of O, Si, Mg and Fe which compose 95\,per cent of the mass of bulk Earth \citep{McDonough2000}. Assuming bulk Earth composition for the parent body and steady-state accretion, we expect these accretion rates to agree. Discrepancies are explained by enhancements or depletions with respect to bulk Earth composition. There is a greater disagreement between the two methods for He-dominated white dwarfs ($\langle\dot{M}_\mathrm{acc, total}-\dot{M}_\mathrm{acc, scaled}\rangle_{\mathrm{He}}=\,0.38\,$dex) compared to H-dominated white dwarfs ($\langle\dot{M}_\mathrm{acc, total}-\dot{M}_\mathrm{acc, scaled}\rangle_{\mathrm{H}}=\,0.33\,$dex) due to the former more likely observed in the decreasing phase where differential sinking results in an apparent enhancement of slow-sinking heavier metals relative to lighter metals. This reduces the usefulness of accretion rates measured from cool He-dominated white dwarfs.

The current understanding of the origin of the accreted material is that a planetary body is scattered towards the white dwarf, then disrupted when entering the the Roche radius forming an eccentric ring of debris which eventually circularises and finally forms a circumstellar disc \citep{Jura2003, Veras2014I, Veras2015, Malamud2021}. Material is then gradually accreted onto the white dwarf, driven by Poynting-Robertson (PR) drag and/or viscous forces in the disc \citep{Rafikov2011, Metzger2012}. The physical  processes involved in the initial disruption and subsequent accretion should be independent of the white dwarf atmospheric composition, and so it is expected that the total accretion rates should be equal for all spectral types. Nevertheless, the left panel of Fig.\,\ref{fig:T_eff_vs_acc_rate} reveals a dichotomy of accretion rates onto H- and He-atmosphere white dwarfs, with median accretion rates of $\dot M_{\text{acc, H}} = 7.7 \times 10^7\,\mathrm{g\,s^{-1}}$ and $\dot M_{\text{acc, He}} = 8.7 \times 10^8\,\mathrm{g\,s^{-1}}$, respectively. This discrepancy has been noted previously \citep{Farihi2012, Bergfors2014}, but there is currently no consensus on its physical origin. For hot white dwarfs ($T_{\mathrm{eff}}\,\gtrsim\,17\,000\,$K), Ca loses its efficacy as a tracer of enrichment since its strength decreases with increasing temperature as \ion{Ca}{II} is ionised to \ion{Ca}{III} \citep{Gaensicke2012}.

Considering PR drag being the primary mechanism for transporting material from the circumstellar disc to the surface of the white dwarf, the maximum accretion rate is $\dot M_{\text{acc}} \simeq 10^8-10^9\,\mathrm{g\,s^{-1}}$ (\citealt{Rafikov2011}; \citealt{Brouwers2022})~--~consistent with the rates onto H-atmospheres, but much lower than those  onto He-atmospheres. Either some aspect of the entire disruption/accretion process onto the white dwarf is not fully understood, or the method of calculating accretion rates is flawed. Since accretion rates are equal to the diffusion rate in the steady-state phase, the accuracy of sinking timescales and the mass of the CVZ could lead to inaccurate results.

Periods of intense accretion could explain the accretion rate dichotomy \citep{Farihi2012}. Because of the short sinking timescales of H-atmospheres (days to $\simeq10^3$\,yr, \citealt{Koester2009_WDs}), short ($\sim1-100$\,yr) but intense bursts during which accretion rate increases by orders of magnitude would be difficult to be detected. For He-atmospheres, the measured accretion rates are time-averages over the sinking timescales of $\simeq10^4-10^6$\,yr, preserving the signature of intense accretion bursts in the form of higher photospheric abundances. At least two possible physical models for burst-like accretion have been put forward. \citet{Brouwers2022} argued that the disruption of large asteroids creates more fragments, leading to more collisions and a rapid grind-down producing dust, causing accretion to occur at a high rate over a short amount of time. Smaller asteroids produce fewer fragments, leading to less collisions producing dust, so accretion occurs over a longer times but at a lower rate. Therefore, they postulate that we observe the ongoing accretion of small asteroids onto H-dominated white dwarfs but can detect the traces of intense accretion events from large asteroids in the atmospheres of He-dominated white dwarfs. Alternatively, \citet{Rafikov2011, Metzger2012} discussed a scenario in which the accretion rate is initially driven by PR, but as dust particles sublimate, gas drag begins to enhance the accretion rate, resulting in a runaway accretion event; \citet{Okuya2023} presented an alternative discussion of the dust-gas interactions resulting in enhanced accretion rates with respect to PR drag. Whether high-accretion-rate bursts are actually happening is unknown, and the above discussion highlights one fundamental shortcoming in the current interpretation of atmospheric abundances, i.e. that the accretion rate onto the white dwarf is constant during the accretion episode.

Another explanation for the dichotomy in the accretion rates is that some of the physics relevant to diffusion and convection within the white dwarf atmospheres and envelopes are not fully understood, and a number of possible extensions to the standard framework have been published. Convective overshoot in DAZs \citep{Cunningham2019} increases the total accretion rate by up to an order of magnitude for white dwarfs with $T_{\text{eff}} < 18\,000$\,K. The X-ray observations of the prototypical metal-enriched G29-38 appear to confirm the predictions of the overshoot model \citep{Cunningham2022_xray}, thus potentially bringing the accretion rates for H-dominated white dwarfs closer, but not quite up to those of He-dominated white dwarfs. Another, even more extreme enhancement of the accretion rates in H-atmosphere white dwarfs is predicted for the case of thermohaline convection  \citep{Deal2013, Wachlin2022, Dwomoh2023}, however, whether this type of convection is physically possible remains debated \citep{Koester2015}. A final note concerning the analysis of H-dominated atmospheres is that radiative levitation becomes increasingly important for $T_\mathrm{eff}\gtrsim18\,000$\,K, adding another systematic uncertainty in the determination of the accretion rates \citep{Koester2014_DAZ}.

In conclusion, measuring the absolute values and the time-dependence of the accretion rates onto metal-enriched white dwarfs remains a difficult endeavour.

\subsection{Minimum mass of accreted bodies}

Accreted material rapidly sinks towards the core of a white dwarf and becomes unobservable, making the mass of an accreted object very difficult to constrain. The absolute minimum mass can be calculated by considering material within the mixing zone of the white dwarf atmosphere. The mixing zone is defined as either the convection zone in cooler white dwarfs or the region above optical depth $\tau_{\text{R}}\,\lesssim\,5$ in the radiative atmosphere of warmer H-atmosphere white dwarfs. The minimum mass, $M_{_{\text{min}}}$, can be calculated using Eq.\,\ref{eq:min_mass} 

\begin{equation}
    M_{_{\text{min}}} = \sum_{_\text{Z}} \text{Z}/\text{H(e)} \frac{\text{A(Z)}}{\text{A(H(e))}} M_{_{\text{MZ}}} = \sum_{_\text{Z}} \tau_{_{\text{Z}}} \dot M_{_{\text{Z}}}
    \label{eq:min_mass}
\end{equation}

\noindent
which is adapted from equations in \citet{Izquierdo2021}, and is independent of the accretion phase. The minimum mass can be computed using knowledge of the photospheric abundance for each metal, Z/H(e), as well as the atomic weight of the element, $A$. $M_{_{\text{min}}}$ can also be calculated by using the accretion rate, $\dot M_{_{\text{Z}}}$, and sinking timescale, $\tau_{_{\text{Z}}}$, of each metal.

The amount of material that has sunk towards the core cannot be determined in any direct way. Efforts have been made to use the measured photospheric number abundances in conjunction with geochemical models to constrain the mass of the accreted body \citep{Buchan2022}, although this method cannot be applied widely as it requires abundances of Ni, Cr, and Si, which are rarely detected together (Fig.\,\ref{fig:T_effhist}).

\begin{figure*}
    \centering
    \includegraphics[width=\textwidth]{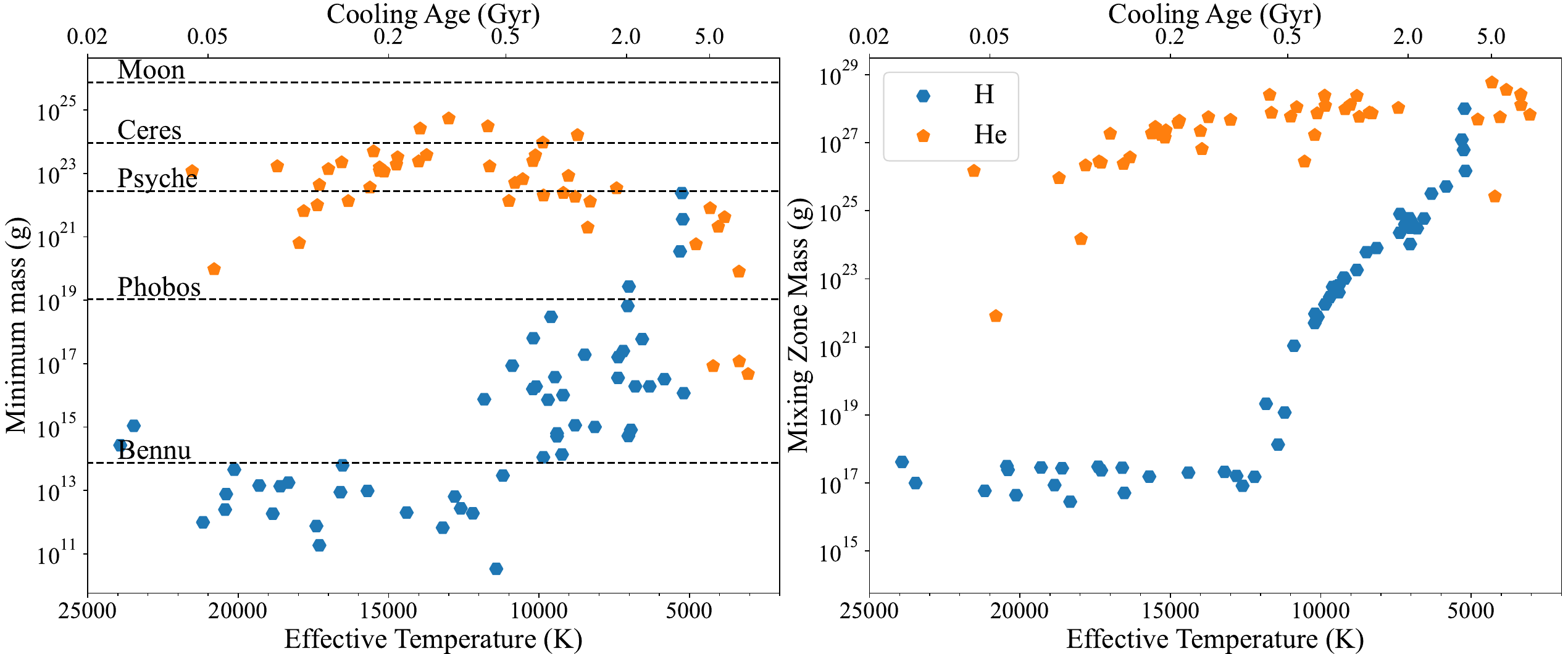}
    \caption{The left panel shows the minimum mass of planetary material in the mixing zones of H- and He-dominated white dwarfs, represented by blue hexagons and orange pentagons respectively. For He-dominated white dwarfs, the mixing zone is defined as the convection zone whereas for warmer H-dominated white dwarfs, it is defined as the region above optical depth $\tau_{\text{R}}\,\lesssim\,5$. Examples of a variety of Solar System objects are included for comparison. The right panel expresses the evolution of the mixing zone mass with decreasing $T_{\text{eff}}$ and increasing cooling age. The much lower mixing zone masses of the two He-dominated white dwarfs with $T_\mathrm{eff}>15\,000$\,K (WD1536+520 \citep{Farihi2016} and WD\,J2047$-$1259 \citep{Hoskin2020}) arise from the large amounts of trace H ($\log(\mathrm{H/He})\simeq-1$ to $-2$) in their atmospheres.}
    \label{fig:min_mass}
\end{figure*}

The minimum masses computed from the data within PEWDD are shown in  Fig.\,\ref{fig:min_mass}, and are consistent with the general assumption that white dwarfs typically accrete asteroid-sized objects \citep{JuraYoung2014}. It is interesting to note that a few He-atmosphere white dwarfs have accreted planetary debris corresponding to the mass of the minor planet Ceres, which alone makes up  25\,per cent of the mass of the Asteroid Belt. Given that the accretion events onto He-atmosphere white dwarfs are only detectable for $\simeq10^4-10^6$\,yr, a very small fraction of their cooling ages, this suggests that the reservoir of planetary bodies feeding white dwarfs is much larger than the Asteroid Belt in the Solar System \citep{Debes2012}. No white dwarf mixing zone contains metals exceeding a Moon mass, suggesting that no rapid accretion (e.g. via direct impact) of moon or planet-sized objects has been detected so far.

The growth of the mixing zone masses with decreasing $T_\mathrm{eff}$ is illustrated in the right panel of Fig.\,\ref{fig:min_mass}. The very low minimum masses for H-dominated white dwarfs with $T_\mathrm{eff}\gtrsim10\,000$\,K arises from the fact that these stars have only shallow convection zones or radiative atmospheres, and provides little physical insight into the architectures of evolved planetary systems. A similar, albeit much weaker trend is seen among the hotter He-dominated white dwarfs.

Comparing the evolution of the minimum accreted masses and mixing zone masses in metal-enriched He-atmospheres reveals an interesting trend: whereas the mixing zone masses gently increase between $20\,000$\,K and 5000\,K, the minimum accreted masses peak around $T_\mathrm{eff}\simeq12\,500$\,K. This may suggest that there is a finite reservoir of large planetary bodies available which becomes depleted within the first $\simeq0.5-1.0$\,Gyr, but future work calculating parent body masses is required to investigate this. The large mixing zone masses, and low opacities of cooler He-atmosphere white dwarfs increases the sensitivity of optical spectra to the presence of photospheric metals, ruling out that the above trend is an observational selection effect. Conversely, at temperatures $>10\,000$\,K, the sensitivity of He-atmospheres for the detection of accretion of small bodies, $M\lesssim10^{19}$\,g drops, and it is hence not clear whether the accretion of such small bodies is common at shorter cooling ages.

\subsection{Discs}

Compact ($\sim$\,Roche-lobe sized) circumstellar discs are detected either by an infrared excess related to the thermal emission of dust \citep{Zuckerman1987, Kilic2006, FarihiDiscs2009} or by metallic emission lines from photo-ionised gas \citep{Gaensicke2006_disc, Gaensicke2007, Gaensicke2008b, Wilson2014, Melis2020}. To date, all white dwarfs hosting a gas disc also exhibit infrared excess. The detection of a dust or gas disc is a strong indication of ongoing accretion.

All metal-enriched H-dominated white dwarfs are very likely to be in the steady-state accretion phase (because of their short sinking timescales), yet not all show an infrared excess, suggesting that some discs have so far escaped detection. Although as many as 25--50\,per cent of white dwarfs show metal-enrichment \citep{Koester2014_DAZ, Zuckerman2003, Zuckerman2010}, only 1--3\,per cent show signs of hosting a dusty disc \citep{FarihiJuraZuckerman2009, Rocchetto2012, TWilson2019}. Even rarer are discs showing gaseous emission, with only 21 such systems reported so far \citep{GentileFusillo2021}, at a rate of just $0.067\,^{+0.042}_{-0.025}$\,per cent \citep{Manser2020}. A clear question therefore emerges: is there any intrinsic difference between these three types of systems?

The canonical model of a dust disc is a flat, optically thick and passive disc which extends from the sublimation radius of $T_{\text{disc}} \approx 1200\,$K to the Roche radius \citep{Jura2003}, set by the temperature and mass of the white dwarf respectively. The disc consists of micron-sized grains which eventually accrete onto the white dwarf surface via PR drag \citep{BochkarevRafikov2011}. The population is more diverse than this simple model, with previous observations finding discs that are cool and truncated \citep{Farihi2008}, potentially optically thin \citep{Bonsor2017}, located beyond the Roche limit \citep{Farihi2017}, or too bright to be accounted for by the canonical model \citep{GentileFusillo2021}.

The distribution of white dwarfs with detected infrared excesses, as a function of the accretion rate onto the white dwarf, is displayed in Fig.\,\ref{fig:T_eff_vs_acc_rate}. Generally, infrared excesses are detected at systems with higher accretion rates \citep{Jura2007}. However, it is important to note that this plot is biased, as systems with large accretion rates and metal-enrichment have been favoured as targets for infrared observations. Nevertheless, \citet{TWilson2019} did carry out a double-blind \textit{HST} and \textit{Spitzer} survey for metal-enrichment and infrared excess, and again found that only $1.5\pm \genfrac{}{}{0pt}{1}{1.5}{0.5}$~per cent of the stars had a dust detection, whereas $45\pm4$~per cent showed traces of metals. The main hypothesis for the non-detection of discs around metal-enriched H-dominated white dwarfs is that the discs may be optically thin, and/or contain a significant fraction of their mass in larger grains, resulting in an overall weaker infrared excess.

There is a notable drop-off in detected infrared excesses at $T_{\text{eff}} < 10\,000\,$K even for white dwarfs with large accretion rates, likely due to detection limits of previous observatories such as \textit{Spitzer} and \textit{WISE}.  An unbiased mid-infrared survey observing H-dominated metal-enriched white dwarfs making use of the much improved sensitivity of \textit{JWST}, compared to previous facilities, would likely reveal the true population of discs. In addition, infrared spectroscopy can identify the mineralogy of discs, with silicates having been detected using \textit{Spitzer} and \textit{JWST} \citep{Jura2007, Swan2024}.

The green diamonds in Fig.\,\ref{fig:T_eff_vs_acc_rate} represent systems where gaseous emission has been detected. There is an absence of gaseous emission around white dwarfs with $T_{\text{eff}} < 12\,000\,$K, which is consistent with the hypothesis that the gas is heated via photo-ionisation~--~cooler white dwarfs have insufficient UV photons to photo-ionise the largely metallic gas \citep{Kinnear2009, Melis2010, GentileFusillo2021}. Gas emission suggests a recent production of gas or a mechanism that generates a constant supply of gas. A puzzling aspect is the location of the gas: whereas irradiation from the white dwarf will result in the total sublimation of the dust in the inner disc, the detected gaseous discs largely overlap with the dust discs, i.e. cover regions where sublimation can not explain the formation of gas \citep{Manser2016}. Hence another mechanism is required to generate the detected gas discs. The highest levels of infrared variability are found among systems that also exhibit gaseous emission \citep{Swan2021}, suggesting that collisions are capable of generating detectable amounts of gas.  Such collisions may arise if sufficiently large, solid bodies are orbiting within the dust disc, perhaps on a somewhat eccentric orbit \citep{Manser2019}. Gas generated by collisions beyond the sublimation radius will re-condense on short time scales \citep{Metzger2012, Okuya2023}. Most gaseous discs have relatively steady emission line fluxes, suggesting ongoing collisions \citep{Manser2016, Dennihy2018, Manser2021}, however, in some cases the gas emission lines have been seen fading away within a decade \citep{Wilson2014}. Finally, there is another class of purely gaseous discs, first identified in \citet{Gaensicke2019} with no dusty component. Rich in volatiles (H, S, O), this disc has been explained by the photo-evaporation of a giant planet atmosphere.

\section{Magnetic fields} \label{sec:magnetic}

Magnetic white dwarfs exhibit Zeeman-split metal lines, which can be used to estimate the strength of the magnetic field \citep[e.g.][]{Angel1974,Jordan1998}. These effects must be included when measuring the photospheric abundances of magnetic white dwarfs \citep[e.g.][]{Farihi2011b,Kawka2021,Hollands2023}. However, for white dwarfs with very strong magnetic fields, $B \gtrsim 100\,$MG (e.g. \citealt{Vennes2024}), synthetic spectra including metals cannot be calculated due to the complex behaviour of atoms in extremely strong fields. For very low magnetic fields, flux spectra do not resolve the Zeeman splitting, and thus spectropolarimetry must be used \citep[e.g.][]{Schmidt1995,LandstreetBagnulo2019}. Here, we investigate whether magnetic metal-enriched white dwarfs differ in any of their properties from their non-magnetic counterparts.

\begin{figure}
    \centering
    \includegraphics[width=\columnwidth]{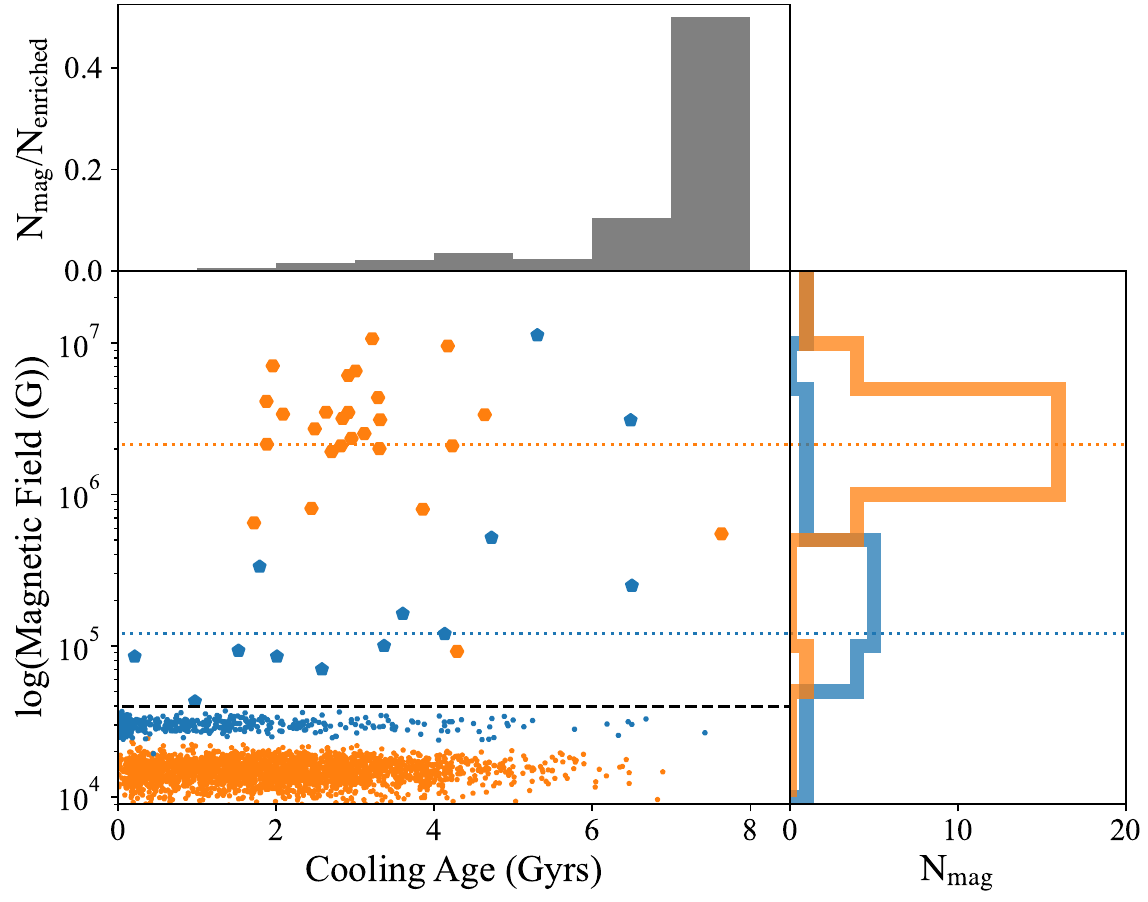}
    \caption{Magnetic field strength as a function of cooling age for metal-enriched H- and He-dominated white dwarfs from PEWDD, in blue pentagons and orange hexagons respectively. The black horizontal dashed line illustrates the threshold for detecting Zeeman splitting in flux spectra. The coloured horizontal dotted lines show the median values of magnetic field strength for H- and He-dominated atmospheres. The upper histogram shows the fraction of magnetic white dwarfs compared to the total number of metal-enriched white dwarfs in PEWDD. The right histogram illustrates the distribution of magnetic field strengths for H- and He-dominated white dwarfs. The points below the detection threshold are white dwarfs where no magnetic field has been detected, scattered to show the cooling age distribution compared to their magnetic counterparts.}
    \label{fig:magvscooling}
\end{figure}

In general, H-dominated white dwarfs have weaker magnetic fields compared to He-dominated white dwarfs, with median magnetic field strengths of $B_{\text{H}} = 0.12\,$MG and $B_{\text{He}} = 2.14\,$MG respectively, as shown in Fig.\,\ref{fig:magvscooling}. Even for white dwarfs with similar cooling ages, there is a clear dichotomy between magnetic fields strength for H-atmosphere and He-atmosphere white dwarfs. Magnetism often only appears in white dwarfs at $\tau_{\text{cool}} \simeq 1-2\,$Gyr \citep{BagnuloLandstreet2022,OBrien2024}, and the very low fraction of younger magnetic metal-enriched white dwarfs (only WD\,2105$-$820; \citealt{Swan2019} and WD\,1350$-$090 \citealt{SchmidtSmith1994,Zuckerman2003} are magnetic and have $\tau_{\text{cool}} < 1\,$Gyr) corroborates this. The onset of magnetism at this cooling age could be caused by the interaction of the growing CVZ with core crystallisation \citep{Isern2017, Schreiber2021}. This hypothesis might also explain metal-enriched H-atmospheres having lower magnetic fields than metal-enriched He-atmospheres, since they generally have smaller CVZs. The origin of weak magnetism in the two young metal-enriched white dwarfs could be remnants of magnetic fields from previous evolutionary stages \citep{BagnuloLandstreet2022}.

As the cooling age increases, there appears to be an increase in the fraction of metal-enriched white dwarfs that have a detectable magnetic field \citep{KawkaVennes2014,Kawka2019,Kawka2021}. There is a gradual increase in the occurrence of magnetism up to $\tau_{\text{cool}} = 4.5\,$Gyr before a dearth of magnetic white dwarfs in the range $4.5 \lesssim \tau_{\text{cool}} \lesssim 6.5\,$Gyr. Finally, magnetic white dwarfs make up a large fraction of the metal-enriched sample for $\tau_{\text{cool}} > 6.5\,$Gyr, although there are much fewer metal-enriched white dwarfs in this range, so the distributions here subject to low number statistics. There is a weak correlation between magnetic field strength and cooling age for He-atmospheres, but this is much more pronounced for H-atmospheres.

It is not well established how magnetic fields affect the likelihood of detecting photospheric metals: sufficiently strong fields will suppress convection in the white dwarf, leading to a radiative atmosphere instead \citep{Wickramasinghe1986,Valyavin2014,Tremblay2015,GentileFusillo2018}. Given that the sinking timescales of radiative atmospheres are significantly shorter than those for their convective counterparts, metals remain detectable for a much shorter period of time. Hence, strong fields may reduce the amount of time for which metal-enrichment can be detected. Finally, for extremely large magnetic fields, metal lines are so smeared that the spectra of these stars become featureless continua \citep{Hollands2015}.

\section{Binaries} \label{sec:binaries}

\citet{Zuckerman2014} investigated the projected separation of metal-enriched white dwarfs in binaries. They found that for separations less than 1000\,au, metal-enrichment is less common, implying that closer companions reduce the formation or long-term stability of planetary systems \citep{Roell2012,Coleman2022,Cuello2023}, resulting in less planetary material to be accreted by the white dwarf. For separations greater than 1000\,au, binaries with metal-enriched and non-enriched white dwarfs are equally common, although this study only contained six metal-enriched white dwarfs. \citet{Noor2024} studied the frequency of DAZ and DZ white dwarfs with wide binary companions compared to field white dwarfs and concludes that the presence of metals in white dwarfs are not the result of planetary system instabilities caused by stellar companions. Their study only made use of the samples of \citet{Dufour2007}, \citet{Hollands2017} and \citet{TWilson2019}, and they find a discrepancy between wide binary fractions for DAZ and DZ white dwarfs.

To enlarge the sample of metal enriched white dwarfs with a wide binary companion, we made use of the astrometric data from \textit{Gaia} in a systematic search for common proper motion companions to all stars in PEWDD using the methods described in \citet{Hollands2018b} and \citet{McCleery2020}. For each metal-enriched white dwarf with a \textit{Gaia} ID that passes the cuts of \verb|parallax_over_error|\,$>\,1$ and \verb|astrometric_excess_noise|\,$<\,1$, we search for objects within a projected separation\footnote{It should be noted that this is a minimum separation, since the unknown inclination reduces the apparent distance between the stars.} of 1\,pc and within a radial distance of 1\,pc. To consider whether the objects share a common proper motion, we calculate the tangential velocity difference using the difference in proper motion. We then apply a cut in projected separation and tangential velocity for objects that could not be gravitationally bound in a binary to remove contaminants.

We identify 40 metal-enriched white dwarfs that fulfil these criteria\footnote{WD\,J091621.31+254029.10 \citep{Hollands2018} is not recovered in the cone search since it is the tertiary member of an hierarchical triple star system. Additionally, SDSS\,J080740.69+493059.7 \citep{Hollands2018} fails the quality cut. Since both are confirmed as binaries in the literature, they are included in Table\,\ref{tab:cone_search}.}, including 8 new systems, shown in Table\,\ref{tab:cone_search}. Four of these wide binaries have hydrogen-dominated atmosphere white dwarf companions, and the remainder have main sequence star companions. These 34 white dwarfs with main sequence star companions are candidates for studies comparing the composition of planetary bodies to stellar abundances, such as in the pilot study of \citet{Bonsor2021}.

\begin{figure}
    \centering
    \includegraphics[width=\columnwidth]{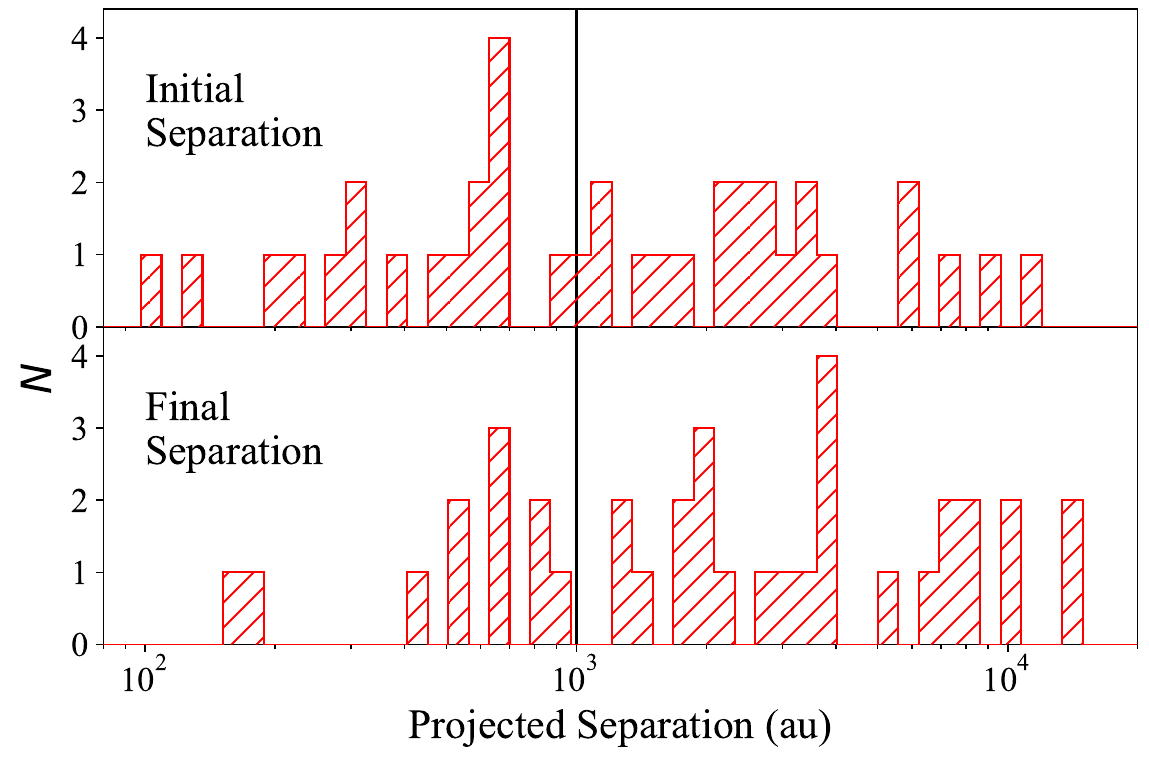}
    \caption{The distribution of metal-enriched white dwarfs in wide binaries from Table\,\ref{tab:cone_search} as a function of projected separation. The top panel shows the estimated initial projected separation been when both stars in the binary were on the main sequence, whereas the bottom panel shows the observed projected separation. The vertical line corresponds to a separation of 1000\,au, which \citet{Zuckerman2014} identified as a threshold below which the frequency of metal-enrichment decreases.}
    \label{fig:binaries}
\end{figure} 

Stars experience mass loss during the red giant phase, which causes the semi-major axis between the two binary companions to increase. A full consideration of this process requires knowledge of the initial and final eccentricities of the systems (e.g. \citealt{Veras2014}; \citealt{Veras2016Review}), which we do not have. For simplicity, we assume adiabatic expansion and that the eccentricity remains constant during the mass loss phase so we can approximate the initial binary semi-major axis, $a_0$ using Eq.\,10 from \citet{ShappeeThompson2013}, knowing the final semi-major axis, $a_f$ and the initial and final total binary mass, $M_0$ and $M_f$ respectively.

\begin{equation}
    a_0 = \frac{a_f M_f}{M_0}
    \label{eq:binary_sep}
\end{equation}

The white dwarf mass is taken from GF21, and the IFMR from \citet{Cummings2018} is used to calculate the initial mass. If the binary companion is a white dwarf, we calculate the initial separation when both stars were on the main sequence. We estimate main sequence star masses by considering the luminosity $L$, calculated from the $G$-band magnitude, $G$, and parallax, $p$, taken from \textit{Gaia} DR3, and the Sun's absolute bolometric magnitude, $M_{\text{bol, \sun}} = 4.74$ \citep{Gaia2016,Gaia2018}.

\begin{equation}
    2.5 \log L = G + 5 - 5\log (1/p) - M_{\text{bol}, \sun}
\end{equation}

The results of this computation is displayed in the top panel of Fig.\,\ref{fig:binaries}, showing the initial projected separation distribution of the binary systems, i.e. when both stars were on the main sequence. Many metal enriched white dwarfs had wide binary companions when on the main sequence.
Recent studies of wide binary companions to planet-hosting main sequence stars suggest that the formation of planetary systems are only significantly suppressed by companions within 50\,au \citep{Kraus2016}, with binary companions at distances greater than 200\,au having negligible effect on planet formation \citep{Moe2021}. Our results agree with the formation of planetary systems suppression within 200\,au, with only two metal enriched white dwarfs with binary companions interior to 200\,au when considering the estimated initial separation. The bottom panel of Fig.\,\ref{fig:binaries} demonstrates that 11 metal-enriched white dwarfs have binary companions within 1000\,au, a substantial increase compared to previous work \citep{Zuckerman2014}. In our larger sample, the frequency of metal white dwarf metal enrichment decreases at a smaller final separation of $\simeq500\,$au.

\section{Conclusions} \label{sec:conclusion}

Metal-enriched white dwarfs allow for the most accurate measurements of exo-planetary bulk compositions. PEWDD is a homogeneous data base of 1739 systems with traces of photospheric metals, designed for statistical studies of the abundances of exo-planetary bodies.

Current studies of metal-enriched white dwarfs are largely limited by observational biases. Metals such as Ca, Cr and Na are much more commonly detected compared to what we would expect assuming bulk Earth abundances. The white dwarf effective temperature plays a major role in  which metals can be detected, as well as the wavelength range of the observations.

We made use of PEWDD to discuss a range of aspects related to metal-enriched white dwarfs, including statistics on the detection of individual elements as a function of $T_\text{eff}$; the dichotomy of H-atmosphere and He-atmosphere white dwarf accretion rates; uncertainties in photospheric abundances; the frequency of magnetism among metal-enriched white dwarfs, and the distribution of their field strengths; and statistics on metal-enrichment among white dwarfs in wide binaries.

PEWDD will be continually updated as the photospheric abundances for more white dwarfs are published. With upcoming large spectroscopic surveys such as DESI, SDSS-V, 4MOST and WEAVE, many more white dwarfs will be inspected for metal-enrichment. As well as further identification of metal-enriched white dwarfs, we emphasise the importance of  more detailed studies of suitable white dwarfs with large numbers of detected elements, as these are most valuable for our insight into the composition of the accreted planetary bodies, and therefore for our understanding of the formation and evolution of exo-planets.

\begin{acknowledgements}
      This research received funding from the European Research Council under the European Union’s Horizon 2020 research and innovation programme numbers 101020057 (JTW, BTG, AS, PI) and 101002408 (MOB). AMC is supported by grant project reference 2590460 from the Science and Technology Facilities Council (STFC). TC is supported by NASA through the NASA Hubble Fellowship grant HST-HF2-51527.001-A awarded by the Space Telescope Science Institute, which is operated by the Association of Universities for Research in Astronomy, Inc., for NASA, under contract NAS5-26555. This research has made use of NASA’s Astrophysics Data System; the SIMBAD database, operated at CDS, Strasbourg, France; the Montreal White Dwarf Database (MWDD; \citealt{MWDD}) and the VizieR service. This work has made use of data from the European Space Agency (ESA) mission Gaia (\href{https://www.cosmos.esa.int/gaia}{https://www.cosmos.esa.int/gaia}), processed by the Gaia Data Processing and Analysis Consortium (DPAC, \href{https://www.cosmos.esa.int/web/gaia/dpac/consortium}{https://www.cosmos.esa.int/web/gaia/dpac/consortium}). Funding for the DPAC has been provided by national institutions, in particular the institutions participating in the Gaia Multilateral Agreement. We thank the anonymous referee for a constructive report.
\end{acknowledgements}

%
%

\bibliographystyle{aa}
\bibliography{bibliography}

\section*{Appendix A: Additional tables}
\begin{appendix}
\renewcommand{\thetable}{A.\arabic{table}}
\setcounter{table}{0}


\begin{table*}
    \caption{Metal-enriched white dwarfs in wide binaries with properties of their companion star.}
    \centering
    \begin{tabular}{{p{0.22\linewidth}|p{0.05\linewidth}|p{0.05\linewidth}|p{0.05\linewidth}|p{0.27\linewidth}|p{0.22\linewidth}}}
    \hline
        WD\,J Name & $T_{\text{eff}}$ & $\log g$ & H/He & Companion star \textit{Gaia} ID & Companion Type \\
        \hline        
        WD\,J003244.21+082720.22 & 7260 & 8.38 & He & Gaia DR3 2749781072426150784 & Main sequence star \\
        WDJ\,004434.77-114836.05 & 5300 & 7.97 & He & Gaia DR3 2377344185944929280 & M-dwarf \citep{Birky2020} \\
        WD\,J004459.06+151821.70 & 6430 & 8.23 & He & Gaia DR3 2781085405419253504 & Main sequence star \\
        WD\,J004521.88+142045.36 & 5030 & 7.89 & H & Gaia DR3 2776464639084019200 & G-type star \citep{Lee1984} \\
        WD\,J010820.77-353441.70 & 28\,800 & 7.86 & H & Gaia DR3 5014008975278023040 & Main sequence star \\ 
        WD\,J011208.58+045508.64 & 20\,090 & 8.23 & He & Gaia DR3 2564020163461415168 & F-type star \citep{Barney1949} \\
        WD\,J013222.84+052925.09 & 7150 & 7.86 & H & Gaia DR3 2564851394251630848 & DA white dwarf \citep{Homeier1998} \\
        WD\,J013557.24-002705.31 & 5570 & 8.03 & He & Gaia DR3 2509739915802134400 & Main sequence star \\
        WD\,J023053.31-475526.11 & 58\,500 & 7.69 & H & Gaia DR3 4939012317940174592 & Main sequence star \\ 
        WD\,J023830.93+063759.84 & 13\,500 & 8.10 & H & Gaia DR3 18493721155296640 & Main sequence star \\
        WD\,J030953.95+150521.83 & 24\,500 & 8.13 & H & Gaia DR3 31047184711544832 & White dwarf \\
        WD\,J051013.94+043838.43 & 20\,600 & 8.01 & H & Gaia DR3 3238868171156736768 & DA white dwarf \citep{Gianninas2011} \\ 
        WD\,J062737.58+100213.87 & 9010 & 8.20 & He & Gaia DR3 3327487537050352640 & M-dwarf \citep{Eggen1965} \\ 
        WD\,J074020.79-172449.16 & 7980 & 8.14 & He & Gaia DR3 5717278911884264576 & M-dwarf \citep{Davison2015} \\ 
        WD\,J080131.13+532900.68 & 8880 & 8.17 & He & Gaia DR3 936641902362276736 & Main sequence star \\
        WD\,J082019.49+253035.54 & 12\,200 & 8.00 & He & Gaia DR3 679451873333368448 & F-type star \citep{Cannon1993} \\ 
        WD\,J082611.70+325000.40 & 6140 & 8.06 & H & Gaia DR3 902973378851513984 & Main sequence star \\
        SDSS\,J080740.69+493059.7 & 5100 & 8.00 & He & Gaia DR3 934631853372231296 & M-dwarf \citep{Hollands2018} \\
        WD\,J091621.31+254029.10 & 5340 & 7.83 & He & Gaia DR3 687914402017314816 & K-dwarf \citep{Stephenson1986} \\
        WD\,J091714.53+263015.31 & 12600 & 8.50 & He & Gaia DR3 694710925769832064 & ?\tablefootmark{a} \\
        WD\,J101201.87-184333.27 & 6390 & 8.18 & He & Gaia DR3 5669429299704388352 & K-dwarf \citep{Upgren1972} \\ 
        WD\,J101955.91+121631.53 & 21\,200 & 7.95 & H & Gaia DR3 3886617649631192320 & DA white dwarf \citep{Andrews2015} \\ 
        WD\,J102251.33+520305.15 & 13000 & 7.91 & He & Gaia DR3 848456534329522560 & Main sequence star \\
        WD\,J103209.99+532935.37 & 44\,500 & 7.88 & H & Gaia DR3 850146823002877440 & Main sequence star \\ 
        WD\,J104156.43+411013.06 & 7280 & 8.21 & He & Gaia DR3 779606421867685504 & Main sequence star \\
        WD\,J112125.75+141713.76 & 10060 & 8.29 & He & Gaia DR3 3966668212166854144 & Main sequence star \\
        WD\,J122650.02+444514.18 & 12190 & 7.97 & He & Gaia DR3 1541711569263907072 & Main sequence star \\
        WD\,J125950.35+043126.42 & 22\,200 & 7.96 & H & Gaia DR3 3705070756418520064 & Main sequence star \\ 
        WD\,J133314.60-675117.19 & 5660 & 8.20 & He & Gaia DR3 5845300239052540416 & Main sequence star \\
        WD\,J142736.17+534828.00 & 14\,600 & 8.07 & He & Gaia DR3 1608497623521965184 & K-dwarf \citep{Xu2017} \\
        WD\,J150119.70+560915.02 & 9320 & 8.03 & He & Gaia DR3 1612720366647934720 & Main sequence star \\
        WD\,J162731.10+485918.83 & 5200 & 8.05 & H & Gaia DR3 1411226580161084160 & Main sequence star \\
        WD\,J163740.64+134036.51 & 6910 & 8.02 & H & Gaia DR3 4461978959028633088 & Main sequence star \\ 
        WD\,J164539.76+305931.71 & 7350 & 8.16 & H & Gaia DR3 1311660510866758016 & Main sequence star \\
        WD\,J194343.68+500437.93 & 33\,920 & 7.97 & H & Gaia DR3 2134964715987301888 & M-dwarf \citep{Gaidos2016} \\
        WD\,J200418.26-560247.93 & 38\,100 & 7.65 & H & Gaia DR3 6448025712768292736 & Main sequence star \\
        WD\,J200940.19-522515.87 & 22480 & 7.90 & He & Gaia DR3 6665910709364013952 & Main sequence star \\
        WD\,J213621.56+113726.87 & 11\,100 & 8.04 & He & Gaia DR3 1766826194114207360 & Main sequence star \\ 
        WD\,J223858.55+131304.24 & 8070 & 8.01 & H & Gaia DR3 2731772377633105024 & Main sequence star \\ 
        WD\,J225435.71+803953.79 & 13\,700 & 8.16 & He & Gaia DR3 2286107290891539840 & Main sequence star \\
        \hline
    \end{tabular}
    \tablefoot{We used $T_{\text{eff}}$ and $\log g$ from GF21. We include stellar type from Simbad, but where this was not available, the object was placed on a Hertzsprung-Russell (HR) diagram to determine whether it is on the main sequence or white dwarf cooling track.\\
    \tablefoottext{a}{Sits below the main sequence but far to the right of the white dwarf cooling track on an HR diagram.}}
    \label{tab:cone_search}
\end{table*}

\begin{table*}
    \centering
    \caption{References for all white dwarfs with entries in PEWDD at time of publication.}
    \begin{tabular}{c|c|c|c}
        \citet{Badenas-Agusti2024} & \citet{Gaensicke2012}  & \citet{Koester1996} & \citet{Swan2023} \\
        \citet{Barstow2014} & \citet{Gaensicke2019}  & \citet{Koester2000} & \citet{Tremblay2020} \\
        \citet{Berger2005} & \citet{Genest-Beaulieu2019}  & \citet{KoesterWilken2006} & \citet{VennesKawka2013} \\
        \citet{Bergeron2011} & \citet{GentileFusillo2017}  & \citet{KoesterKepler2015} & \citet{Vennes2010} \\
        \citet{Bergfors2014} & \citet{GonzalezEgea2021}  & \citet{Koester1997} & \citet{Vennes2011} \\
        \citet{Blouin2018a} & \citet{Grenfell1974} & \citet{Koester2005a} & \citet{Wegner1973} \\
        \citet{Blouin2018b} & \citet{Holberg1996} & \citet{Koester2005b} & \citet{WehrseLiebert1980} \\
        \citet{Blouin2019} & \citet{Hollands2017} & \citet{Koester2014_DAZ} & \citet{Wilson2014} \\
        \citet{Blouin2020} & \citet{Hollands2021} & \citet{Koester2014b} & \citet{Wilson2015} \\
        \citet{Blouin2022} & \citet{Hollands2022} & \citet{Leggett2018} & \citet{Wilson2019} \\
        \citet{CottrellGreenstein1980} & \citet{Hoskin2020} & \citet{Liebert1977} & \citet{Wolff2000} \\
        \citet{Coutu2019} & \citet{Izquierdo2021} & \citet{Liebert1987} & \citet{Wolff2000} \\
        \citet{Dennihy2016} & \citet{Johnson2022} & \citet{Manser2016} & \citet{Xu2013} \\
        \citet{Desharnais2008} & \citet{JuraXu2010} & \citet{MelisDufour2017} & \citet{Xu2014} \\
        \citet{Dobbie2005} & \citet{Jura2009} & \citet{Melis2012} & \citet{Xu2017} \\
        \citet{Doyle2023} & \citet{Jura2012} & \citet{Melis2018} & \citet{Xu2019} \\
        \citet{Dufour2007} & \citet{Jura2015} & \citet{Melis2020} & \citet{Zeidler1986} \\
        \citet{Dufour2010} & \citet{Kaiser2021} & \citet{OBrien2023} & \citet{Zuckerman1998} \\
        \citet{Dufour2012} & \citet{KawkaVennes2011} & \citet{OBrien2024} & \citet{Zuckerman2003} \\
        \citet{Dupuis1993} & \citet{KawkaVennes2012} & \citet{Owens2023} & \citet{Zuckerman2007} \\
        \citet{Elms2022} & \citet{KawkaVennes2014} & \citet{Raddi2015} & \citet{Zuckerman2010} \\
        \citet{Farihi2010} & \citet{Kawka2011} & \citet{Rogers2024b} & \citet{Zuckerman2011} \\
        \citet{Farihi2011b} & \citet{Kawka2019} & \citet{Rogers2024} & \citet{Zuckerman2013} \\
        \citet{Farihi2011a} & \citet{Kawka2021} & \citet{Sayres2012} \\
        \citet{Farihi2012} & \citet{Kenyon1988} & \citet{ShipmanGreenstein1983} \\
        \citet{Farihi2013} & \citet{Kepler2015} & \citet{Shipman1977} \\
        \citet{Farihi2016} & \citet{Kepler2016} & \citet{Sion1984} \\
        \citet{FortinArchambault2020} & \citet{Kepler2021} & \citet{Sion1988} \\
        \citet{Friedrich1999} & \citet{Kilic2020} & \citet{Steele2021} \\
        \citet{Friedrich2000} & \citet{Klein2010} & \citet{Subasavage2007} \\
        \citet{Gaensicke2007} & \citet{Klein2011} & \citet{Subasavage2017} \\
        \citet{Gaensicke2008b} & \citet{Klein2021} & \citet{Swan2019} \\
    \end{tabular}
    \label{tab:database_references}
\end{table*}

\end{appendix}

\end{document}